\documentclass{aa}
\usepackage{hyperref}
\usepackage{lineno}
\nolinenumbers
\usepackage{natbib,twoopt}
\bibpunct{(}{)}{;}{a}{}{,}    
\usepackage{graphicx}
\usepackage{txfonts}
\usepackage{lscape}
\usepackage{parskip}
\usepackage{lipsum}
\usepackage{url}

\def\sci{Science}
\def\araa{Annu. Rev. Astron. Astrophys.}

\newcommand\nata{Nature Astronomy}

\def\fesc{\ifmmode f_{\rm esc} \else $f_{\rm esc}$\fi}

\begin{document}

\title{A great diversity of spectral shapes in the ionising spectra of
$z$ $\sim$ 0.6 -- 1 galaxies revealed by \textit{HST}/COS and possible detection
of nebular LyC emission}

\titlerunning{A great diversity of spectral shapes in the ionizing spectra of
$z$ $\sim$ 0.6 -- 1 galaxies}

\author{Y. I. Izotov\inst{1},
D. Schaerer\inst{2,3},
G. Worseck\inst{4}, 
N. G. Guseva\inst{1},
A. Verhamme\inst{2},
C. Simmonds\inst{5,6},
J. Chisholm\inst{7} 
          }

\institute{Bogolyubov Institute for Theoretical Physics,
National Academy of Sciences of Ukraine, 14-b Metrolohichna str., Kyiv,
03143, Ukraine\\
\email{yizotov@bitp.kyiv.ua}
\and
Observatoire de Gen\`eve, Universit\'e de Gen\`eve, 
51 Ch. des Maillettes, 1290, Versoix, Switzerland
\and
IRAP/CNRS, 14, Av. E. Belin, 31400 Toulouse, France
\and
17VDI/VDE Innovation+Technik, Berlin, Germany
\and
The Kavli Institute for Cosmology (KICC), University of Cambridge, Madingley Road, Cambridge CB3 0HA, UK
\and
Cavendish Laboratory, University of Cambridge, 19 JJ Thomson Avenue, Cambridge CB3 0HE, UK
\and
Astronomy Department, University of Texas at Austin,
2515 Speedway, Stop C1400 Austin, TX 78712-1205, USA
}

\date{Accepted XXX. Received YYY; in original form ZZZ}

\abstract{We present observations of eleven compact star-forming
galaxies in the redshift range $z$ = 0.6145 -- 1.0053, with the Cosmic Origins Spectrograph (COS) on board the \textit{Hubble Space Telescope} (\textit{HST}). We aim to spectroscopically measure for the first time the Lyman continuum (LyC) over a wider  
rest-frame wavelength range of $\sim$ 600 -- 900\AA\ compared to
$\sim$ 850 -- 900\AA\ in previous studies of galaxies at $z$ $\sim$ 0.3 -- 0.4.
The \textit{HST} data are supplemented by SDSS spectra 
of all galaxies and by a VLT/Xshooter spectrum of one galaxy, J0232$+$0025. These data are used to derive the spectral energy distribution in the entire 
UV and optical range, the stellar mass, and the chemical composition from 
the nebular emission lines. We detect stellar LyC emission in seven out of
eleven galaxies with escape fractions, $f_\textrm{esc}$(LyC), in the range of
$\sim$ 2 -- 60\%, and establish upper
limits for $f_\textrm{esc}$(LyC) in the remaining galaxies.
We discover for the first time nebular LyC emission  
as a bump just bluewards of the LyC limit at 912\AA\ in two galaxies, 
J0232$+$0025 and J1021$+$0436. We find a similar bump among our earlier
studies in a less distant galaxy J1243$+$4646 with $z$ = 0.4317. We conclude
that the use of the LyC continuum in the wavelength range close to the LyC
limit, which contains both the stellar and nebular continua, requires special 
consideration to not overestimate the observed $f_\textrm{esc}$(LyC).}


   \maketitle

\section{Introduction}\label{intro}

Star-forming galaxies (SFGs) are considered important contributors to the reionisation of the Universe at redshift $z$ $\ga$ 6. It has been suggested that, at these redshifts, SFGs were responsible for the major part of the ionising radiation \citep*{O09,WC09,M13,Y11,B15a,Fi19,Le20,Me20,T20,Ch22}. However, to achieve this,
their ionising photon production and the escape fraction of the Lyman continuum (LyC) should be $\xi_\textrm{ion}$~$\sim$~10$^{25.2}$~Hz~erg$^{-1}$ and $\gtrsim 10$\%, respectively \citep[e.g. ][]{O09,R13,D15,Robertson15,K16}.
On the other hand, the role of active galactic nuclei (AGN)
in reionisation is less clear \citep[e.g. ][]{Fi19,M24}.

The observations of escaping hydrogen ionising emission in galaxies at $z\gtrsim 4$ are virtually impossible because of the faintness of the galaxies and the high opacity of the partially ionised intergalactic medium \citep{In14,Wo14,R22}.
The LyC escape fraction can be sufficient for reionisation, provided that the SFGs at $z\gtrsim 6$ responsible for reionisation are similar to these lower-redshift galaxies \citep{Va15,B16,Sh16,B17,Va18,RT19,Sa20,Vi20,Ma17,Ma18,St18,Be22,Gr22,Sa22,Ki23,Liu23,D24,Ke24}.

  \begin{table*}
  \caption{Coordinates, redshifts, distances, and integrated characteristics of selected galaxies \label{tab1}}
\begin{tabular}{lrrccrcrccr} \hline
Name&R.A.(2000.0)&Dec.(2000.0)&$z$&$D_L$$^\textrm{a}$&\multicolumn{1}{c}{$D_A$$^\textrm{b}$}&12+logO/H&O$_{32}$&log($M_\star$/M$_\odot$)$^\textrm{c}$&$M_\textrm{FUV}^\textrm{d}$&SFR$^\textrm{e}$ \\ \hline
J0232+0025&02:32:29.50&+00:25:04.28&0.7641&4803&1543&8.13&5.8& 9.41&$-$21.37& 45 \\
J0256+0122&02:56:58.26&+01:22:45.80&0.8867&5773&1622&7.95&2.8& 8.84&$-$21.62& 53 \\
J0815+2942&08:15:53.71&+29:42:29.40&0.9141&5994&1636&8.36&1.9&10.01\,\,\,&$-$21.26& 34 \\
J0837+4512&08:37:58.69&+45:12:33.28&0.7816&4939&1556&7.93&3.6& 8.18&$-$21.23&157 \\ 
J0901+5111&09:01:38.28&+51:11:46.58&0.6374&3844&1434&7.99&6.4& 9.20&$-$20.12& 29 \\
J0908+4626&09:08:25.75&+46:26:50.46&0.8148&5199&1579&8.16&4.0& 9.55&$-$21.18& 53 \\
J0955+3935&09:55:28.37&+39:35:53.11&0.8293&5314&1588&7.83&2.7&10.09\,\,\,&$-$21.03& 44 \\
J1021+0436&10:21:40.54&+04:36:47.80&1.0053&6745&1677&8.42&5.1&10.67\,\,\,&$-$22.64&126 \\
J1252+5237&12:52:31.39&+52:37:16.41&0.6225&3735&1419&7.98&4.3& 8.43&$-$21.38& 41 \\ 
J1358+4611&13:58:06.75&+46:11:43.19&0.6195&3713&1416&8.06&2.8& 9.22&$-$21.65& 51 \\ 
J1450+3913&14:50:29.88&+39:13:21.83&0.6145&3676&1410&8.11&3.5& 9.85&$-$20.98& 35 \\ 
\hline
\end{tabular}
\tablefoot{$^\textrm{a}$Luminosity distance in Mpc \citep[NED, ][]{W06}. $^\textrm{b}$Angular size distance in Mpc \citep[NED, ][]{W06}. $^\textrm{c}$Stellar mass $M_\star$ in solar units. $^\textrm{d}$Rest-frame absolute magnitude at $\lambda$=1500\AA\ derived from an SED fitting of the SDSS spectrum and an extrapolation of the SED to the UV range. $^\textrm{e}$Star-formation rate in units of M$_\odot$ yr$^{-1}$ derived from the extinction-corrected H$\beta$ luminosity.}
  \end{table*}

Low-mass compact SFGs at $z$~$\la$~1 are considered local proxies of high-redshift LyC leaking galaxies. Many of these galaxies at $z$ $\sim$ 0.1 -- 0.3 are named ‘Green Pea’ (GP) galaxies because of their intense green colour in SDSS images \citep{Ca09}, while others are classified as ‘luminous compact galaxies’ (LCG) over a wider redshift range $z$ = 0.02 -- 0.63, selected from both the SDSS images and spectra \citep{I11}. Indeed, \citet{I15} and \citet{B23} compared various properties of high-redshift SFGs with those of low-redshift compact SFGs and found that they are very similar. In particular, stellar masses, star formation rates (SFR), and O$_{32}$ = [O~\textsc{iii}]$\lambda$5007/[O~\textsc{ii}]$\lambda$3727 ratios of compact SFGs overlap with those of high-redshift SFGs in wide ranges of parameters. Recent \textit{James Webb Space Telescope} (\textit{JWST}) spectroscopic observations of $z$ $\sim$ 8 SFGs have revealed that their chemical composition is similar to that of local SFGs \citep[e.g. ][]{S22}. \citet{W21}, \citet{Rh23}, \citet{Mat23}, \citet{Ha24}, and \citet{AC25} have also shown that high-redshift galaxies have properties similar to those of GPs at $z$~$\la$~0.3.

\textit{HST}/COS observations of a large number of compact SFGs at redshifts $z\sim$ 0.3 -- 0.5 in a wide range of stellar masses and O$_{32}$ ratios have been obtained by \citet{I16a,I16b,I18a,I18b,I21a,I22}, \citet{MM21}, \citet{Fl22a,Fl22b}, and \citet{Xu22}. According to the authors, many of these galaxies leak LyC emission, with an escape fraction of up to 72~\%. However, we note that all measurements of LyC emission in galaxies at $z$=0.3--0.5 were done near the Lyman limit, within a narrow rest-frame wavelength range between 800 -- 900\AA\ (only part of which is usable due to residual geocoronal emission in the \textit{HST} spectra). In this range, the contribution from nebular emission -- originating from the recombination of the free electron on the hydrogen ground state -- and contamination by the geocoronal Ly$\alpha$ and O~\textsc{i} $\lambda$1300 emission lines can be important, potentially causing the derived $f_\textrm{esc}$(LyC) to be overestimated \citep{In10,Si24,Wo14,W16,Ma21}.
We take a huge effort to minimise geocoronal contamination (scattered light), for example by comparing with orbital night data. Geocoronal emission sets usable wavelength ranges in the LyC, depending on the redshift.
The \textit{HST} observations of galaxies at lower redshifts, $z$ $<$ 0.1, \citep{L16,K24} also indicate that they are LyC leaking objects. Furthermore, the Ly$\alpha$ profiles in the spectra of most of these galaxies have double-peaked profiles with a small velocity separation between the peaks, indicating low H~\textsc{i} column densities and therefore high escape fractions
\citep{V15,Hen18,Ch20,Ga20,I16a,I16b,I18a,I18b,I21a,I22}.

\begin{figure*}
\centering
\includegraphics[angle=0,width=0.75\linewidth]{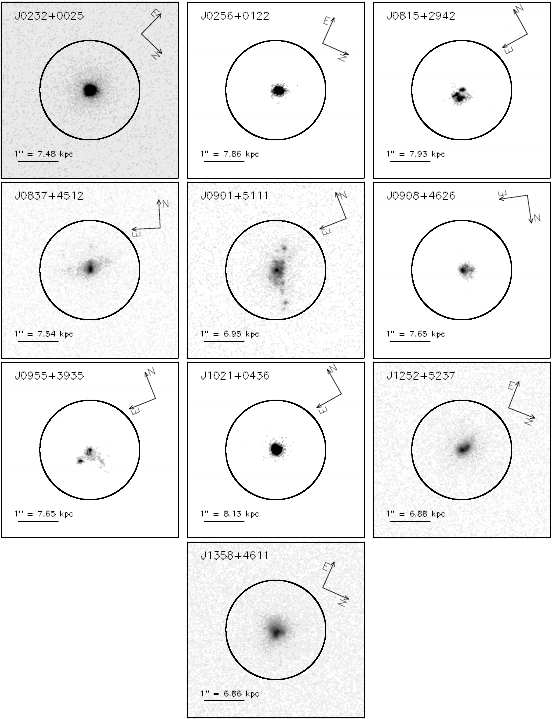}
\caption{\textit{HST} COS/NUV acquisition images of the candidate LyC leaking galaxies on a surface brightness scale. The COS spectroscopic aperture with a diameter of 2\farcs5 is indicated in all panels by a circle. The linear scale in each panel is derived by adopting an angular size distance (Table~\ref{tab1}).
\label{fig1}}
\end{figure*}

In this paper, we extend the redshift range of LyC leakers observed with the
\textit{HST}/COS to $z$~$\sim$~0.6~--~1.0. The goal of these observations is two-fold: 1) to determine the LyC escape fraction $f_\textrm{esc}$(LyC) from the measurement of the LyC emission in the wavelength range free of nebular emission, as well as from a measurement that includes both stellar and nebular emission; and 2) to study the shape of both the stellar and nebular LyC continua in galaxies with detected emission in a large range of wavelengths from $\sim$~600~--~900\AA\ (to be addressed in a forthcoming paper). To this end, we analyse new \textit{HST}/COS observations of the LyC in 11 compact SFGs. The selection criteria are presented in Section~\ref{sec:select}. The \textit{HST}/COS and \textit{HST}/STIS observations and data reduction are described in Sections~\ref{sec:obs} and \ref{sec:obs1}. The surface brightness (SB) profiles in the NUV band are discussed in Section~\ref{sec:sbp}. The properties of the galaxies derived from the observations in the optical range are considered in Section~\ref{sec:ext}. The comparison of the \textit{HST}/COS spectra with the extrapolation of the modelled SEDs to the UV range is made in Section~\ref{sec:global}. The escaping LyC emission is discussed in Section~\ref{sec:lyc} together with the corresponding escape fractions. The likely detection of nebular LyC emission in two galaxies is presented in Section~\ref{nebular}. The indirect indicators of escaping LyC emission are considered in Section~\ref{Ind}. We summarise our findings in Section~\ref{summary}.

To derive absolute magnitudes and other integrated parameters, we adopted luminosity and angular size distances \citep{W06} with the cosmological parameters $H_0$=67.1 km s$^{-1}$Mpc$^{-1}$, $\Omega_\Lambda$=0.682, and $\Omega_m$=0.318 \citep{P14}. These distances are presented in Table~\ref{tab1}.

\section{Selection criteria}\label{sec:select}

The eleven galaxies in the redshift range $z = 0.6145 - 1.0053$ for observations with the \textit{HST} are selected from the SDSS data release 16 \citep{A20}. They satisfy the following conditions: 
1) The hydrogen emission lines H$\delta$ $\lambda$4102, H$\gamma$ $\lambda$4340, H$\beta$ $\lambda$4861, and the oxygen emission lines [O~\textsc{iii}] $\lambda$4959, 5007\AA\ are strong, and the [O~\textsc{iii}] $\lambda$4363 line is detected; these are used to determine oxygen abundance, expressed as 12+log(O/H).
In particular, the equivalent width of the [O~\textsc{iii}] $\lambda$5007 emission line is above 400\AA\ , and this line is detected with a signal-to-noise ratio (S/N)
above 20. We note that the H$\alpha$ $\lambda$6563 emission line is not observed because it is outside the wavelength range of the SDSS spectrum.
2) The selected galaxies are the brightest compact SFGs in the adopted redshift range, have the highest O$_{32}$ ratios, and have the highest equivalent widths EW(H$\beta$) of the H$\beta$ emission line. This condition
is required in order for the LyC to be bright enough to detect. These galaxies are listed in Table \ref{tab1}. Despite the larger redshifts, they have properties similar to those of LyC leakers at lower redshifts \citep{I16a,I16b,I18a,I18b,I20,I21a,I22,Fl22a}.

However, at variance with previous lower-redshift observations, the redshift of the selected galaxies is too high to observe the Ly$\alpha$ emission line with the COS far ultraviolet (FUV)
medium-resolution gratings, precluding the possibility to resolve its profile. The angular full widths at half maximum (FWHM) of the selected galaxies in the SDSS $g$-band images range from 1\farcs1 -- 1\farcs6, which is much smaller than the 2\farcs5 diameter COS aperture, allowing the detection of all galaxy UV emission. The SDSS and apparent \textit{GALEX} magnitudes of the selected galaxies are shown in Table \ref{taba1}.

  \begin{table}
  \caption{\textit{HST}/COS observations \label{tab2}}
  \begin{tabular}{lccc} \hline
\multicolumn{1}{c}{}&\multicolumn{1}{c}{}&\multicolumn{2}{c}{Exposure time (s)} \\ 
\multicolumn{1}{c}{Name}&\multicolumn{1}{c}{Date}&\multicolumn{2}{c}{(Central wavelength (\AA))} \\ 
    &    &MIRRORA&G140L\\ \hline
J0232+0025&2021-09-24&2$\times$1338     &10742\\
            &          &         &(800)\\
J0256+0122&2023-08-01&2$\times$300      &12288\\
            &          &         &(800)\\
J0815+2942&2023-04-02&2$\times$300      &9950$^\textrm{b}$\\
            &          &         &(800)\\
J0837+4512&2022-02-11&2$\times$1409     &11315\\
            &          &         &(800)\\
J0901+5111&2022-04-15&2$\times$1433     &11507\\
            &          &         &(800)\\ 
J0908+4626&2024-09-24&2$\times$300      &10159 \\
            &          &         &(800)\\
J0955+3935&2024-04-08&2$\times$300      &7346$^\textrm{b}$\\
            &          &         &(800)\\
J1021+0436&2024-04-07&2$\times$300      &1799$^\textrm{b}$ \\
          &2024-12-03&2$\times$0$^\textrm{a}$      &7683$^\textrm{b}$ \\
            &          &         &(800)\\
J1252+5237&2022-02-20&2$\times$1433     &11507\\
            &          &         &(800)\\ 
J1358+4611&2022-02-18&2$\times$1409     &11315\\
            &          &         &(800)\\ 
J1450+3913&2021-06-11&2$\times$0$^\textrm{a}$       &  8233$^\textrm{b}$\\
            &          &         &(800)\\
\hline
\end{tabular}
\tablefoot{$^\textrm{a}$Failed exposure. $^\textrm{b}$Shortened exposure due to technical failure.}
\vspace{0.5cm}
\caption{\textit{HST}/STIS observations\label{tab3}}
  \begin{tabular}{lccc} \hline
\multicolumn{1}{c}{}&\multicolumn{1}{c}{}&\multicolumn{2}{c}{Exposure time (s)} \\ 
\multicolumn{1}{c}{Name}&\multicolumn{1}{c}{Date}&\multicolumn{2}{c}{(Central wavelength (\AA))} \\ 
    &    &MIRVIS&G230L\\ \hline
J0232+0025&2023-08-25&2$\times$200     &10125\\
            &          &         &(2375)\\
\hline
\end{tabular}

  \end{table}

\section{Observations}\label{sec:ob}

\subsection{\textit{HST}/COS observations}\label{sec:obs}

\textit{HST}/COS spectroscopy of the selected galaxies was obtained in programmes
GO~16271 and GO~17171 (PI: D.~Schaerer) during the periods June 2021 -- April
2022 and April 2023 -- December 2024, respectively (Table~\ref{tab2}). The
galaxies were acquired by COS near ultraviolet (NUV) imaging. The brightest NUV
region of each target was centred in the $\simeq 2\farcs5$ diameter
spectroscopic aperture (Fig.~\ref{fig1}). The \textit{HST} acquisition sequence
failed for J1450$+$3913 and for the second visit of J1021$+$0436
(Table~\ref{tab2}), possibly degrading their spectrophotometric quality due to
the vignetting of the COS aperture.

The spectra were obtained with the grating G140L in the 800\,\AA\ setup
(wavelength range 770--1950\,\AA, covering the redshifted LyC emission for
all targets). The spectra of J0232$+$0025 and J1450$+$3913 were obtained at COS
Lifetime Position 4 (resolving power $R\simeq 1250$ at 1500\,\AA\ for a point
source or compact galaxy), while the rest of the spectra were recorded at COS
Lifetime Position 3, where the different light path yields a higher resolution
($R\simeq 2400$ at 1500\,\AA). Typically, the G140L sub-exposures were taken at
all four focal-plane offset positions to correct for COS detector blemishes
except for the observations of four targets (Table~\ref{tab2}), which suffered
from COS (re-)acquisition failures (i.e. shortened exposure time). However,
the quality of the co-added spectra is still sufficient for our main scientific
purpose.

After completion of the \textit{HST} programmes the individual exposures were
reduced with contemporary versions of the \textsc{calcos} pipeline (v3.3.10
for GO~16271 and v3.4.4 for GO~17171) and associated calibration files, followed
by accurate background subtraction and co-addition with \textsc{faintcos}
\citep{Ma21}. We used the same methods as, for example in \citet{I18a,I18b}, but chose
slightly different extraction apertures (GO~16271: 25 pixels and GO~17171:
29 pixels) to preserve the accuracy of the flux calibration at the longest
wavelengths for the highest-redshift targets. Moreover, due to infrequent
monitoring of the COS dark current at the COS detector voltage used for G140L
observations for GO~17171, we used a large base set of 139 dark exposures taken
during the whole observing period and covering a range of orbital conditions to
estimate the dark current in the extraction aperture with \textsc{faintcos}.
By doing so, we assumed that the COS detector suffers negligible gain sag in
the extraction aperture over the observing period (21 months). Still, the
minor adjustments in data processing between the two \textit{HST}/COS programmes
do not significantly change our results. The accuracy of our custom correction
for scattered light in the COS G140L data was verified by comparing the LyC fluxes
obtained in total exposure and in orbital night. We find good
agreement between the two sets of measurements.

Statistical Poisson uncertainties for the reduced spectra were determined
using the approach by \citet{FC98}, as described in \citet{W16} and
\citet{Ma21}. We also included estimates of the systematic error stemming from
background uncertainties using Monte Carlo simulations, as described in
\citet{W16} and \citet{Ma21}. The reason for this is that \citet{FC98} assume
the background is known (i.e. background uncertainty is zero), which is not the
case for our data. 
For the measured average LyC fluxes in 20\,\AA\ bins, we maximised the Poisson
likelihood function \citep[e.g. Equation 2 in][]{Ma21} and computed
statistical uncertainties again using the frequentist method of \citet{FC98}.
The systematic error due to background uncertainties was estimated with
Monte Carlo simulations \citep[as in][]{W16,Ma21}. Thus, the approach is the
same as for individual pixels of the spectra but accounts for pixel-to-pixel
variations in the flatfield, pixel exposure time, and background.
In general, systematic errors are small.

\subsection{\textit{HST}/STIS observations}\label{sec:obs1}
Additionally, for the galaxy J0232$+$0025 we used public \textit{HST}/STIS
observations with the low-resolution G230L grating (programme GO~17169,
PI: D.\ Schaerer) processed with the \textit{HST} \textsc{calstis} pipeline.
Details of these observations are available in
Table~\ref{tab3}. The obtained spectrum covers the rest-frame wavelength range
from $\sim$ 1000 -- 1880\AA. Only a few spectral features are detected in this
spectrum, including hydrogen Ly$\alpha$ $\lambda$1216\AA\ in emission,
Ly$\beta$ $\lambda$1026\AA\ in absorption, C~\textsc{iv} $\lambda$1549\AA\
exhibiting a P-Cygni profile, and, possibly, a weak stellar
O~\textsc{vi} $\lambda$1035\AA\ line with a likely P-Cygni profile.

\subsection{Results from rest-frame optical spectra}\label{sec:Xshooter}

Spectral observations of J0232+0025 in the wavelength range 3000 -- 24000\AA\
were carried out with the Xshooter spectrograph mounted at the VLT/UT2
in nodding-on-slit mode during several nights in 2021
(ESO Program ID 0105.20LV.001, PI:~D. Schaerer).
\textsc{iraf} was used to reduce the observations, which included
subtraction of the bias exposures in the ultraviolet and blue (UVB) and visual
(VIS) arms and of the dark exposures in
the near-infrared (NIR) arm. Applying the \textit{crmedian} routine, we removed the cosmic rays.
Corrections for pixel sensitivity, background subtraction, wavelength
calibration, as well as distortion and tilt of each frame were performed. The
1D spectra were extracted from 2D frames in apertures of 1\farcs6
along the slit for the UVB and VIS arms and 2\farcs4 for the NIR arm.
The flux-calibrated rest-frame spectrum is shown in Fig.~\ref{figa1}.
Emission-line fluxes were measured using total
integral fluxes with the \textsc{iraf} \textit{splot} routine
(Table~\ref{taba2}). For all sources we also measured the line properties from
the SDSS spectra using the \textsc{iraf} {\it splot} routine (Table~\ref{taba3}).

\section{Surface brightness distribution in the NUV range}\label{sec:sbp}

Our galaxies in the COS NUV acquisition images generally have a complex
structure and consist of several bright compact regions of star formation, along
with an extended low-surface-brightness component. The nature of the extended
emission is unclear. However, we note that part of the emission
is diffuse Ly$\alpha$ emission, which is more extended than the stellar
light \citep{Fl22b,Sa25}. This emission line in our galaxies
falls within the NUV range. The SB profiles of our galaxies are
derived, using the COS 
NUV acquisition images and the routine \textit{ellipse} in
\textsc{iraf}/\textsc{stsdas}. The profiles were scaled to magnitudes per square
arcsec, using the apparent \textit{GALEX} NUV magnitudes. There is no
SB profile for the galaxy J1450$+$3913 because its acquisition exposure failed,
as noted above. 

As found previously by \citet{I16a,I16b,I18a,I18b,I20,I21a,I22}, the outer parts
of our galaxies are characterised by a linear weakening in SB
(in magnitudes per square arcsecond), characteristic of disc structures, 
while the central part containing bright star-forming region(s) shows a sharp
increase in J0232+0025, J0256+0122, J0908+4626, and J1021+0436, which have
compact structures. In contrast, a more modest increase is present in the
remaining galaxies, which exhibit complex and extended morphology
(Figs.~\ref{fig1} and \ref{figa2}).

The scale lengths $\alpha$ of our galaxies, defined in Eq.~1 of  \citet{I16b},
range from $\sim$ 0.5 -- 1.0 kpc (Fig.~\ref{figa2}). This is somewhat lower than
that in other LyC leakers at lower redshifts with
$M_\star$/M$_\odot$ $\sim$ 10$^9$ -- 10$^{10}$ \citep{I16a,I16b,I18a,I18b} but is
higher than that in LyC leakers with lower stellar masses \citep{I21a,I22}.
The radii $r_{50}$ $\sim$ 0.5 kpc of our galaxies (Fig.~\ref{figa2}) and
stellar masses are similar to that of galaxies from the LzLCS+ sample
\citep{Fl22a}, which includes all objects observed and analysed earlier by
\citet{I16a,I16b,I18a,I18b,I21a}.

  \begin{table*}
  \caption{Electron temperatures, electron number densities, and 
element abundances in H~\textsc{ii} regions derived from SDSS spectra \label{tab4}}
  \begin{tabular}{lcccccc} \hline
Galaxy &J0232$+$0025 &J0256$+$0122 &J0815$+$2942 &J0837$+$4512 &J0901$+$5111 &J0908$+$4626  \\ \hline
$T_\textrm{e}$ ($[$O \textsc{iii}$]$), K      & 12250$\pm$2420       & 13570$\pm$2450       & 10550$\pm$2700     & 14080$\pm$3380& 14670$\pm$1730 & 11210$\pm$2450       \\
$T_\textrm{e}$ ($[$O \textsc{ii}$]$), K       & 12050$\pm$2240       & 13080$\pm$2210       & 10530$\pm$2500     & 13430$\pm$3010 & 13810$\pm$1520& 11120$\pm$2310        \\
$N_\textrm{e}$ ($[$S \textsc{ii}$]$), cm$^{-3}$&   100$\pm$10       &   100$\pm$10  &        100$\pm$10   &    100$\pm$10  &      100$\pm$10 &      100$\pm$10      \\ 
12+log O/H                             &8.13$\pm$0.06        &7.95$\pm$0.04        &8.36$\pm$0.38      &7.93$\pm$0.05  &7.99$\pm$0.02&8.13$\pm$0.08          \\ 
log Ne/O                               &~$-$0.73$\pm$0.16~~\, &~$-$0.69$\pm$0.12~~\, &~$-$0.90$\pm$0.40~~\,&~$-$0.71$\pm$0.15~~\, &~$-$0.85$\pm$0.08~~\, &~$-$0.65$\pm$0.20~~\,    \\ 
log Mg/O                               &~$-$1.10$\pm$0.21~~\, &~$-$1.33$\pm$0.17~~\, &~$-$1.35$\pm$0.63~~\, &~$-$1.64$\pm$0.27~~\, &~$-$1.39$\pm$0.12~~\, &~$-$1.24$\pm$0.28~~\,   \\
\hline
Galaxy &J0955$+$3935 &J1021$+$0436 &J1252$+$5237 &J1358$+$4611 &J1450$+$3913  \\ \cline{1-6}
$T_\textrm{e}$ ($[$O \textsc{iii}$]$), K      & 15420$\pm$2670       & 10240$\pm$3150       & 13770$\pm$2020& 12490$\pm$1670 & 12440$\pm$1640       \\
$T_\textrm{e}$ ($[$O \textsc{ii}$]$), K       & 14230$\pm$2300       & 10190$\pm$3020       & 13220$\pm$1810 & 12250$\pm$1540& 12210$\pm$1520        \\
$N_\textrm{e}$ ($[$S \textsc{ii}$]$), cm$^{-3}$&   100$\pm$10       &   100$\pm$10  &    100$\pm$10  &      100$\pm$10 &      100$\pm$10      \\ 
12+log O/H                             &7.83$\pm$0.03        &8.41$\pm$0.15        &7.98$\pm$0.03  &8.06$\pm$0.04&8.04$\pm$0.04          \\ 
log Ne/O                               &~$-$0.83$\pm$0.10~~\, &~$-$0.76$\pm$0.33~~\, &~$-$0.74$\pm$0.10~~\, &~$-$0.75$\pm$0.11~~\, &~$-$0.83$\pm$0.11~~\,    \\ 
log Mg/O                               &~$-$1.43$\pm$0.14~~\, &~$-$1.30$\pm$0.48~~\, &~$-$1.06$\pm$0.14~~\, &~$-$1.29$\pm$0.14~~\, &~$-$1.59$\pm$0.15~~\,   \\
\cline{1-6}
\end{tabular}

  \end{table*}

\begin{table}
\centering
  \caption{Electron temperatures, electron number density, and 
element abundances obtained from the Xshooter spectrum of J0232+0025 \label{tab5}}
  \begin{tabular}{lc} \hline
Property & Value \\ \hline
$T_\textrm{e}$ ($[$O \textsc{iii}$]$), K      & 13150$\pm$240 \\
$T_\textrm{e}$ ($[$O \textsc{ii}$]$), K       & 12780$\pm$220 \\
$T_\textrm{e}$ ($[$S \textsc{iii}$]$), K      & 12360$\pm$200 \\
$N_\textrm{e}$ ($[$S \textsc{ii}$]$), cm$^{-3}$&  202$\pm$149 \\ 
12+log O/H                             &8.11$\pm$0.02\\ 
log N/O                                &~$-$1.07$\pm$0.04~~\, \\ 
log Ne/O                               &~$-$0.76$\pm$0.04~~\, \\ 
log Mg/O                               &~$-$1.23$\pm$0.03~~\, \\ 
log S/O                               &~$-$1.72$\pm$0.03~~\, \\ 
log Ar/O                               &~$-$2.58$\pm$0.04~~\, \\ 
log C/O                               &~$-$0.72$\pm$0.04~~\, \\ 
\hline
\end{tabular}

  \end{table}

\section{Properties of galaxies from optical spectra}\label{sec:ext}

\subsection{Interstellar extinction}

Interstellar extinction was derived from the observed decrement 
of all available hydrogen emission lines, which are measurable in the SDSS 
spectra of all galaxies and in the Xshooter spectrum of J0232$+$0025 
\citep{ITL94}. The emission-line fluxes were corrected for the internal 
extinction of the galaxies by adopting the \citet{C89} reddening law 
with $R(V)$~=~3.1 and $A(V)$~=~3.1$\times$$E(B-V)$, 
where $E(B-V)$~=~$C$(H$\beta$)/1.47 \citep{A84}. The fluxes of hydrogen 
emission lines were corrected for underlying stellar absorption. 
Differential extinction $E(B-V)$ = 0.0 -- 0.15 in our galaxies is low 
and is in the range of values obtained by \citet{I16a,I16b,I18a,I18b,I21a} and \citet{Ch22} 
for LyC leaking galaxies. We note that the H$\alpha$ 
emission line is observed only in the Xshooter spectrum of J0232+0025 
(Table~\ref{taba2}), whereas this line is outside the wavelength range of the SDSS 
spectra because of the high redshift (Table~\ref{taba3}).
The extinction-corrected emission line fluxes were used to derive ionic and total 
element abundances following the methods described in \citet{I06} and 
\citet{G13}. 

The emission-line fluxes $I$($\lambda$)/$I$(H$\beta$) corrected 
for extinction, rest-frame equivalent widths (EW),
extinction coefficient 
$C$(H$\beta$), and observed H$\beta$ fluxes in the SDSS spectra are 
shown in Table \ref{taba3}. Corresponding quantities for J0232+0025 obtained from the 
Xshooter spectrum are shown in Table \ref{taba2}.

\subsection{Physical conditions and element abundances}

The fluxes and the direct $T_\textrm{e}$ method were used to derive 
the electron temperature, the electron number density, and 
the abundances of O, Ne, and Mg from the SDSS spectra and those of O, C, N, Ne, Mg, S, and
Ar from the Xshooter spectrum.
These quantities for the SDSS spectra are shown in Table \ref{tab4}; corresponding quantities for the
Xshooter spectrum of J0232+0025 are shown in Table \ref{tab5}.
The electron temperatures $T_\textrm{e}$(O~\textsc{iii}) and
$T_\textrm{e}$(O~\textsc{ii}) derived from the Xshooter spectrum of J0232$+$0025 are
higher than in the SDSS spectrum, but are consistent within the errors.
We note that the errors in the electron temperatures derived from the SDSS
spectra are high, reaching up to $\sim$~30~\% of their values
(Table~\ref{tab4}). This results in relatively high errors in the element
abundances. On the other hand, the error in the electron temperature derived
from the Xshooter spectrum of J0232$+$0025 is much lower (Table~\ref{tab5}).
We also note that the SDSS spectra do not cover the
wavelength range containing 
[S~\textsc{ii}] $\lambda$$\lambda$6717,6731\AA\ emission lines due to the high redshift. 
Fortunately, the electron temperature and element abundances are almost
independent of $N_\textrm{e}$ if the number densities are $\la$ 10$^4$ cm$^{-3}$, which
is expected in our galaxies. Despite these differences and assumptions,
the agreement between the 12~+~logO/H, log Ne/O and log Mg/O from the SDSS and
Xshooter spectra of J0232$+$0025 are good.
The oxygen abundances of the galaxies studied here are 
comparable to those in known low-redshift LyC leakers by
\citet{I16a,I16b,I18a,I18b} and \citet{Fl22a}.
The ratios of the $\alpha$-element (neon, magnesium, sulphur, and argon)
abundances to oxygen abundance are similar to those in
dwarf emission-line galaxies \citep[e.g. ][]{I06,G13,AC24,Z25,Es25}.
The carbon-to-oxygen
abundance ratio in J0232+0025 is similar to that in low-redshift SFGs \citep[e.g. ][]{IT99,Be16,Be19,I23}, whereas the 
nitrogen-to-oxygen abundance ratio in J0232+0025 is somewhat elevated
compared to that of nearby SFGs \citep[e.g. ][]{AC25}, but
similar to that in other LyC leakers at $z$~$\ga$~0.3
\citep[e.g. ][]{I23}. It is derived from the
Xshooter spectrum. No carbon and nitrogen abundances are available from SDSS
spectra for this and other galaxies, since the C~\textsc{iii}] $\lambda$1909 and
[N~\textsc{ii}] $\lambda$6584 emission lines are outside the wavelength range of the
SDSS spectra.

\subsection{Absolute magnitudes, H$\beta$ luminosities, and star formation rates}
\label{sec:absmag}

We derived absolute magnitudes from the fluxes
of the extinction-corrected optical spectral energy distribution (SED),
extrapolated to the rest-frame wavelength of $\lambda$ = 1500 \AA. We designate
these absolute magnitudes as $M_\textrm{FUV}$ (Table~\ref{tab1}).

The H$\beta$ luminosity $L$(H$\beta$) and the corresponding
SFRs were obtained from the extinction-corrected H$\beta$ 
fluxes, using the relation of \citet{K98} for the SFR. 
Here $f_\textrm{esc}$(LyC)
was derived from the stellar LyC fluxes in the regions where the contribution
of nebular LyC emission is low, 
and we discuss in Sects.~\ref{sec:lyc} and \ref{nebular} how $f_\textrm{esc}$ depends on this choice.
The SFRs were multiplied by a factor of 1/[1--$f_\textrm{esc}$(LyC)] to correct
for escaping LyC radiation and are shown in Table~\ref{tab1}. This factor
holds for the H~\textsc{ii} region with an isotropic distribution of the neutral gas; however,
it may be considerably lower if LyC escape is anisotropic. 
The derived SFRs are several times higher than SFRs
for other LyC leakers studied by \citet{I16a,I16b,I18a,I18b,I21a,I22} and
\citet{Fl22a,Fl22b}.

\subsection{Spectral energy distributions and stellar masses}\label{sec:sed}

We used the SDSS spectra of our LyC leakers to fit the SED and derive their 
stellar masses. The fitting method, using a two-component model, is described 
for example in \citet{I18a,I18b}.
The star formation history was approximated by a young instantaneous 
burst with a randomly varying age $t_b$ ($<$ 10 Myr), and by continuous star formation at older ages between $t_1$ and $t_2$
($t_1$ $<$ $t_2$), both randomly varying within the range 10 Myr - 10 Gyr. For
stellar populations, we adopted \textsc{starburst99} models \citep{L99}.

The emission of the nebular continuum at $\lambda$ $>$ 1000\AA\ \citep{A84},
including free-free and free-bound hydrogen
and helium emission, and two-photon emission, was also accounted for 
using the observed H$\beta$ flux, as well as the temperature and density of the H~\textsc{ii}
region. The fraction of nebular emission in the observed continuum near H$\beta$
was determined from the ratio of the observed H$\beta$ equivalent width 
EW(H$\beta$)$_\textrm{obs}$, reduced to the rest-frame, to the equivalent
EW(H$\beta$)$_\textrm{rec}$ = $\epsilon$(H$\beta$)$_\textrm{rec}$/$\epsilon_\textrm{ff}$(4861)
for pure nebular emission, where $\epsilon$(H$\beta$)$_\textrm{rec}$ and
$\epsilon_\textrm{ff}$(4861) are emissivities of the recombination H$\beta$ emission
line and the monochromatic free-free emission at the wavelength of 4861\AA, respectively.
The EW(H$\beta$)$_\textrm{rec}$ does not depend on the electron number density and is
only slightly dependent on the electron temperature, varying from
$\sim$1100\AA\ at $T_\textrm{e}$ = 10000K to $\sim$900\AA\ at
$T_\textrm{e}$ = 20000K \citep{A84}. Recently, \cite{Sc25} obtained similar
values of EW(H$\beta$)$_\textrm{rec}$, ranging from $\sim$1200\AA\ at
$T_\textrm{e}$ = 10000K to $\sim$800\AA\ at $T_\textrm{e}$ = 20000K.
A $\chi^2$ minimisation technique was used firstly to fit the continuum 
in the wavelength range 2500 -- 6000\AA, and secondly to reproduce 
the observed H$\beta$ equivalent width.

The stellar (red lines), nebular (blue lines), and total (black lines) SEDs
superposed on the extinction-corrected rest-frame spectra (grey lines) are
shown in Fig.~\ref{figa3}. The total SEDs reproduce energy distributions in
observed spectra at $\lambda$ $>$ 3000\AA. On the other hand, the observed
spectra at shorter wavelengths in five of the eleven galaxies lie above the
modelled total SEDs, likely due to issues with the flux calibration of the
SDSS spectra.

\section{Comparison of the \textit{HST}/COS spectra with the modelled
SEDs in the UV range}\label{sec:global}

To derive the fraction of escaping ionising radiation, we used two methods based on a comparison between the observed flux in the LyC range and the intrinsic flux produced by stellar populations in the galaxy. 
The intrinsic LyC flux can be obtained either from SED fitting of the SDSS spectra or from the flux of the H$\beta$ emission line.
To verify the quality of our SED fitting, we extrapolate the attenuated SEDs to the UV range and compare them with the observed COS spectra in Fig.~\ref{figa4}. For comparison, we also show with blue-filled circles the \textit{GALEX} FUV and NUV fluxes, along with the fluxes in the SDSS $u,g,r,i,z$ bands.

We find that the spectroscopic and photometric data in the optical range are consistent, indicating that almost all the continuum emission of our galaxies is inside the spectroscopic aperture. Therefore, aperture corrections are not needed. 
On the other hand, considerable deviations of the attenuated SED extrapolations to brighter values from the observed COS spectra are found for the galaxy J1450$+$3913, which had a failed acquisition exposure.

Compared to the SDSS spectrum of J0232$+$0025, the Xshooter spectrum of this galaxy has a significantly higher signal-to-noise ratio and a wider wavelength coverage, including the H$\alpha$ emission line, which is redshifted out of the SDSS bandpass for all of the studied galaxies. This allows a better determination of emission line fluxes, extinction, and the SED. These data are compared with the corresponding data obtained from the SDSS spectrum of J0232$+$0025.

\begin{figure*}
\centering
\includegraphics[angle=0,width=0.70\linewidth]{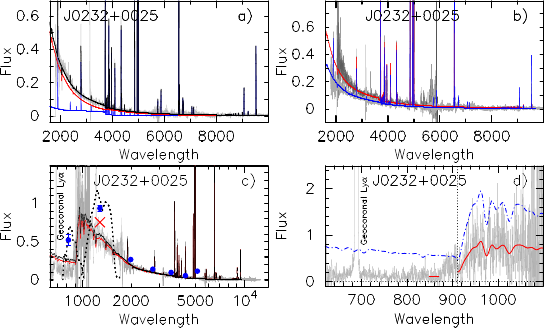}
\caption{(a) Extinction-corrected rest-frame Xshooter spectrum of J0232+0025 (grey line) overlaid with modelled SEDs. Nebular, stellar, and total (nebular+stellar) SEDs are represented by blue, red, and black lines, respectively. (b) Observed rest-frame SDSS (black line) and Xshooter (grey line) spectra of J0232+0025 overlaid with their corresponding attenuated modelled SEDs (the SDSS spectrum shown in red and the Xshooter spectrum shown in blue). (c) Observed rest-frame COS, STIS, and Xshooter spectra of J0232+0025 (grey lines) overlaid with modelled SEDs attenuated by adopting either $R(V)$ = 3.1 (black line) and $R(V)$ = 2.7 (red line), assuming the \citet{C89} reddening law. Blue-filled circles denote \textit{GALEX} FUV and NUV, and SDSS $u,g,r,i,z$ photometric data. The red cross indicates the NUV flux calculated from the SED by adding the Ly$\alpha$ emission with an EW(Ly$\alpha$) = 108\AA\ from the STIS spectrum. Dotted black lines indicate \textit{GALEX} FUV and NUV response curves reduced to the J0232+0025 rest-frame. (d) Observed COS spectrum of J0232+0025 (grey line) overlaid with the intrinsic modelled SED (dash-dotted blue line) and the SED attenuated with $R(V)$ = 2.7. The short horizontal red line represents the level of the LyC stellar continuum, and the vertical dotted line indicates the location of the LyC limit at 912\AA.
\label{fig2}}
\end{figure*}

In Fig.~\ref{fig2}a, we show stellar (red line), nebular (blue line), and total (black line) SEDs obtained from the extinction-corrected Xshooter spectrum of J0232$+$0025 (grey line). As shown, the total SED (black line) reproduces the observed spectrum (grey line) very well.

In Fig.~\ref{fig2}b we compare the observed Xshooter (grey line) and SDSS (black line) spectra with their corresponding total SEDs (red and blue lines). The agreement between the two spectra is good at $\lambda$~$\ga$~4000\AA. However, at shorter wavelengths, the SDSS spectrum lies significantly above the Xshooter spectrum, indicating a problem with the flux calibration of the SDSS spectrum. Consequently, we use the Xshooter spectrum.

\begin{figure}
\centering
\includegraphics[angle=0,width=0.75\linewidth]{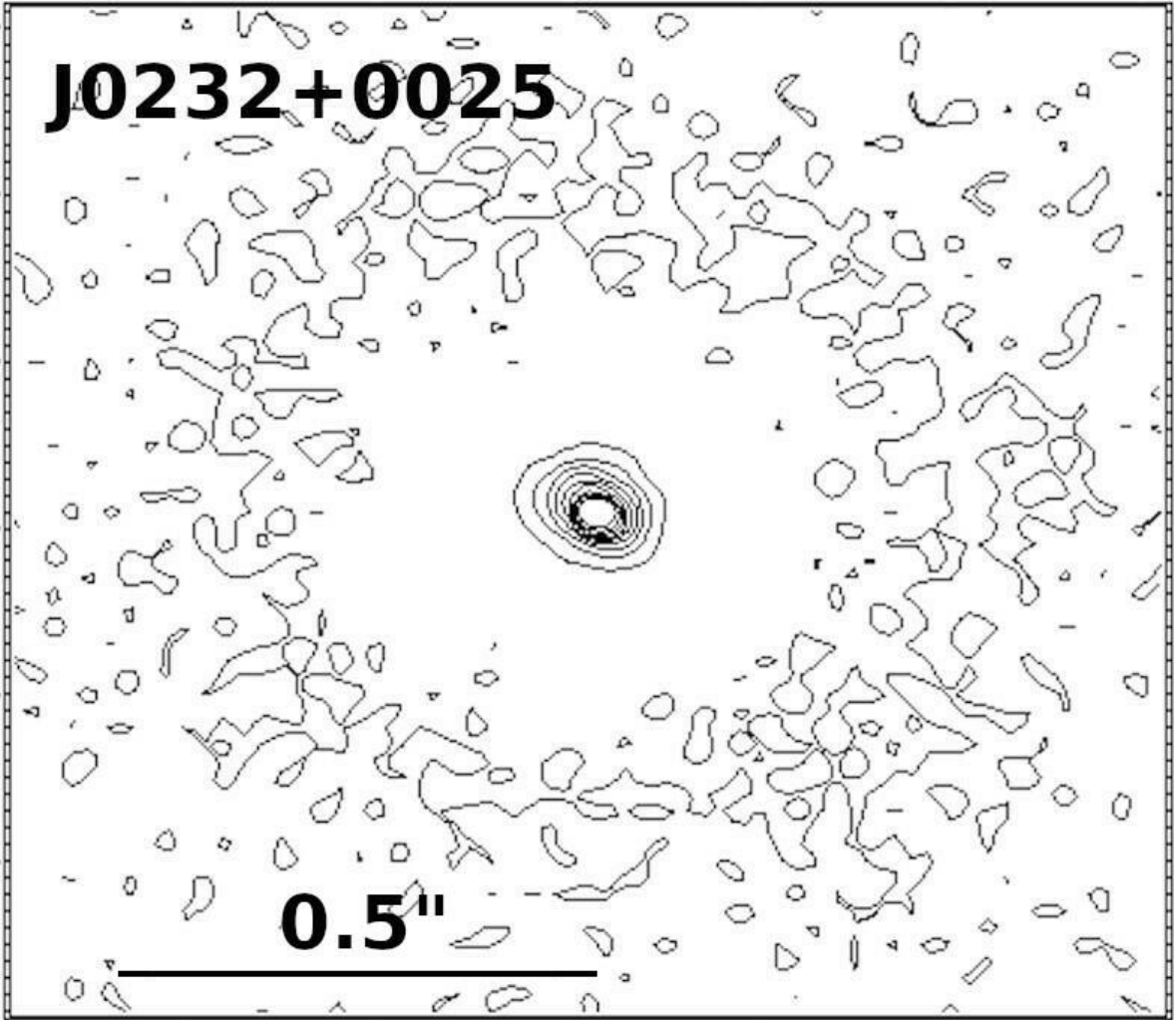}
\caption{Isophotes of the J0232+0025 COS/NUV acquisition image in arbitrary units but with equal steps. The diameter of the outer isophote is larger than the slit width of 0\farcs5 used in the STIS observations.
\label{fig3}}
\end{figure}

In Fig.~\ref{fig2}c we compare the attenuated SED and its extrapolation to the UV range (red and black lines for $R(V)$ values of 3.1 and 2.7, respectively) for J0232$+$0025 with the COS spectrum in the UV range and the Xshooter spectrum in the optical range (grey lines). We find very good agreement between the modelled SEDs and the observed spectra. This contrasts with the large deviation of the SED in the UV range obtained from the SDSS spectrum and the COS spectrum (Fig.~\ref{fig2}a). We also note a large offset between the \textit{GALEX} NUV flux and the SED. This may be partly due to uncertainties in \textit{GALEX} NUV photometry, although the error bars in Fig.~\ref{fig2}c are small.
We suspect that the difference between \textit{GALEX} and SED fluxes in the NUV range is due to the contribution of the Ly$\alpha$ emission line to the \textit{GALEX} NUV flux. Due to the large redshift of the galaxy, this line is located outside the observed COS spectrum but coincides with the position of maximum transmission of the \textit{GALEX} NUV filter (dotted line). Adopting an equivalent width measured in the STIS spectrum, EW(Ly$\alpha$) = 108\AA, we obtain an NUV Ly$\alpha$ plus SED flux (shown by a red cross in Fig.~\ref{fig2}c) that is considerably lower than the observed \textit{GALEX} NUV flux (blue-filled circle). This difference cannot be explained by the small uncertainties in the \textit{GALEX} NUV magnitude (Table~\ref{taba1}). We suggest that the difference is caused by the presence of the extended Ly$\alpha$ halo in J0232$+$0025, with a radius several times larger than that of the stellar emission, similar to haloes
observed in other SFGs \citep[e.g. ][]{C21,E23,R23,T25}. The angular diameter of the region with likely Ly$\alpha$ emission in J0232$+$0025, of at least 1\arcsec\ corresponding to a linear diameter of 7.5 kpc, is considerably larger than the slit width of 0\farcs5 used during STIS observations, whereas most of the stellar emission in the NUV band lies within the STIS slit (Fig.~\ref{fig3}).

\begin{table*}
\centering
  \caption{LyC escape fraction \label{tab6}}
\begin{tabular}{lcccrrrr} \hline
Name&$\lambda_0$$^\textrm{a}$&$A$(LyC)$_\textrm{MW}$$^\textrm{b}$&$I_\textrm{mod}$$^\textrm{c,d}$&\multicolumn{1}{c}{$I_\textrm{obs}$$^\textrm{c,e}$}&\multicolumn{1}{c}{$I_\textrm{esc}$$^\textrm{c,f}$}&\multicolumn{1}{c}{$f_\textrm{esc}$$^\textrm{g}$}&\multicolumn{1}{c}{$f_\textrm{esc}$$^\textrm{h}$} \\
    &(\AA)&(mag)&&&&\multicolumn{1}{c}{(\%)}&\multicolumn{1}{c}{(\%)} \\
\hline
\multicolumn{8}{c}{SDSS+COS} \\ \vspace{0.1cm}
J0232$+$0025$^\textrm{i}$&880-912&0.159& 93.84$\pm$1.92&31.29$^{+2.07}_{-2.16}$$^{+0.06}_{-0.00}$    &36.27$^{+2.40}_{-2.50}$$^{+0.07}_{-0.00}$&38.66$^{+2.56}_{-2.67}$$^{+0.07}_{-0.00}$&47.65$^{+3.15}_{-3.29}$$^{+0.09}_{-0.00}$ \\ \vspace{0.1cm}
J0232$+$0025$^\textrm{i}$&850-870&0.159& 93.84$\pm$1.92& 8.56$^{+1.53}_{-1.46}$$^{+0.00}_{-0.02}$    & 9.85$^{+1.76}_{-1.68}$$^{+0.00}_{-0.02}$&10.55$^{+1.89}_{-1.81}$$^{+0.00}_{-0.02}$&12.94$^{+2.31}_{-2.21}$$^{+0.00}_{-0.03}$ \\ \vspace{0.1cm}
J0256$+$0122&850-890&0.510& 39.53$\pm$3.95& $<$1.70$^\textrm{j}$&$<$2.76&$<$7.01&$<$4.46 \\ \vspace{0.1cm}
J0815$+$2942&850-890&0.214& 57.49$\pm$3.29& $<$1.80$^\textrm{j}$&$<$2.21&$<$3.82&$<$6.21 \\ \vspace{0.1cm}
J0837$+$4512&870-890&0.162& 60.28$\pm$3.70& 2.12$^{+1.25}_{-1.27}$$^{+0.09}_{-0.08}$& 2.42$^{+1.43}_{-1.45}$$^{+0.10}_{-0.09}$& 4.03$^{+2.38}_{-2.41}$$^{+0.17}_{-0.15}$& 2.52$^{+1.49}_{-1.51}$$^{+0.11}_{-0.10}$ \\ \vspace{0.1cm}
J0901$+$5111&870-890&0.137& 43.73$\pm$4.87& $<$1.10$^\textrm{j}$& $<$1.20     &   $<$2.70   & $<$1.60 \\ \vspace{0.1cm}
J0908$+$4626&850-890&0.103& 55.46$\pm$4.50& 2.57$^{+4.36}_{-2.08}$$^{+0.00}_{-0.49}$&2.77$^{+4.70}_{-2.25}$$^{+0.00}_{-0.53}$&4.94$^{+8.38}_{-4.00}$$^{+0.00}_{-0.94}$&3.56$^{+6.04}_{-2.88}$$^{+0.00}_{-0.68}$ \\ \vspace{0.1cm}
J0955$+$3935&850-890&0.074& 58.22$\pm$5.82& 1.86$^{+1.59}_{-1.33}$$^{+0.19}_{-0.43}$    & 1.97$^{+1.68}_{-1.41}$$^{+0.20}_{-0.46}$& 3.39$^{+2.90}_{-2.42}$$^{+0.35}_{-0.78}$ & 3.17$^{+2.71}_{-2.27}$$^{+0.32}_{-0.73}$ \\ \vspace{0.1cm}
J1021$+$0436$^\textrm{i}$&820-912&0.134&119.41$\pm$3.41&41.18$^{+3.20}_{-2.86}$$^{+0.04}_{-0.34}$    &46.56$^{+3.62}_{-3.23}$$^{+0.05}_{-0.38}$&38.98$^{+3.03}_{-2.71}$$^{+0.04}_{-0.32}$ &38.49$^{+3.00}_{-2.67}$$^{+0.04}_{-0.32}$ \\ \vspace{0.1cm}
J1021$+$0436$^\textrm{i}$&750-800&0.134&119.41$\pm$3.41& $<$0.14$^\textrm{j}$& $<$0.16     &   $<$0.13   &    $<$0.13     \\ \vspace{0.1cm}
J1252$+$5237&870-890&0.072& 62.98$\pm$5.30& 4.06$^{+1.21}_{-1.12}$$^{+0.00}_{-0.03}$    & 4.36$^{+1.30}_{-1.20}$$^{+0.00}_{-0.03}$& 6.93$^{+2.07}_{-1.91}$$^{+0.00}_{-0.05}$& 3.47$^{+1.03}_{-0.96}$$^{+0.00}_{-0.03}$ \\ \vspace{0.1cm}
J1358$+$4611&870-890&0.067&108.99$\pm$3.02& 2.52$^{+1.29}_{-1.21}$$^{+0.00}_{-0.03}$    & 2.72$^{+1.39}_{-1.31}$$^{+0.00}_{-0.03}$& 2.52$^{+1.29}_{-1.21}$$^{+0.00}_{-0.03}$& 1.81$^{+0.93}_{-0.87}$$^{+0.00}_{-0.02}$ \\  \vspace{0.1cm}
J1450$+$3913&870-890&0.067& 87.29$\pm$6.51& $<$0.92$^\textrm{j}$ & $<$1.02     &   $<$1.12   & $<$1.02 \\ 
\multicolumn{8}{c}{XShooter+COS} \\ \vspace{0.1cm}
J0232$+$0025&880-912&0.159& 56.01$\pm$0.52&31.29$^{+2.07}_{-2.16}$$^{+0.06}_{-0.00}$    &35.98$^{+2.38}_{-2.48}$$^{+0.07}_{-0.00}$&64.40$^{+4.26}_{-4.45}$$^{+0.12}_{-0.00}$&37.51$^{+2.48}_{-2.59}$$^{+0.07}_{-0.00}$ \\ \vspace{0.1cm}
J0232$+$0025&850-870&0.159& 56.01$\pm$0.52& 8.56$^{+1.53}_{-1.46}$$^{+0.00}_{-0.02}$    & 9.85$^{+1.76}_{-1.68}$$^{+0.00}_{-0.02}$&17.62$^{+3.15}_{-3.01}$$^{+0.00}_{-0.04}$&10.25$^{+1.83}_{-1.75}$$^{+0.00}_{-0.02}$ \\
\hline
  \end{tabular}
\tablefoot{$^\textrm{a}$Rest-frame wavelength range in angstrom used to determine the LyC flux. $^\textrm{b}$Milky Way extinction at the mean observed wavelengths of the range used to determine the LyC flux. The \citet{C89} reddening law with $R(V)$ = 3.1 is adopted here. $^\textrm{c}$Quantities
expressed in 10$^{-18}$ erg s$^{-1}$cm$^{-2}$\AA$^{-1}$. $^\textrm{d}$Intrinsic LyC flux derived from the modelled SED. $^\textrm{e}$Observed LyC flux derived from the spectrum with shadow exposure. $^\textrm{f}$LyC flux corrected for the Milky Way extinction. $^\textrm{g}$$f_\textrm{esc}$(LyC) = $I_\textrm{esc}$(total)/$I_\textrm{mod}$, with $I_\textrm{mod}$ derived from the modelled SED (first method). $^\textrm{h}$$f_\textrm{esc}$(LyC) = $I_\textrm{esc}$(total)/$I_\textrm{mod}$, with $I_\textrm{mod}$ derived from H$\beta$ flux (second method). $^\textrm{i}$Two wavelength ranges considered for the determination of $f_\textrm{esc}$(LyC). The first corresponds to the LyC bump, and the second to the stellar LyC emission. $^\textrm{j}$1$\sigma$ confidence upper limit.}
\end{table*}

\begin{figure*}
\centering
\includegraphics[angle=0,width=0.75\linewidth]{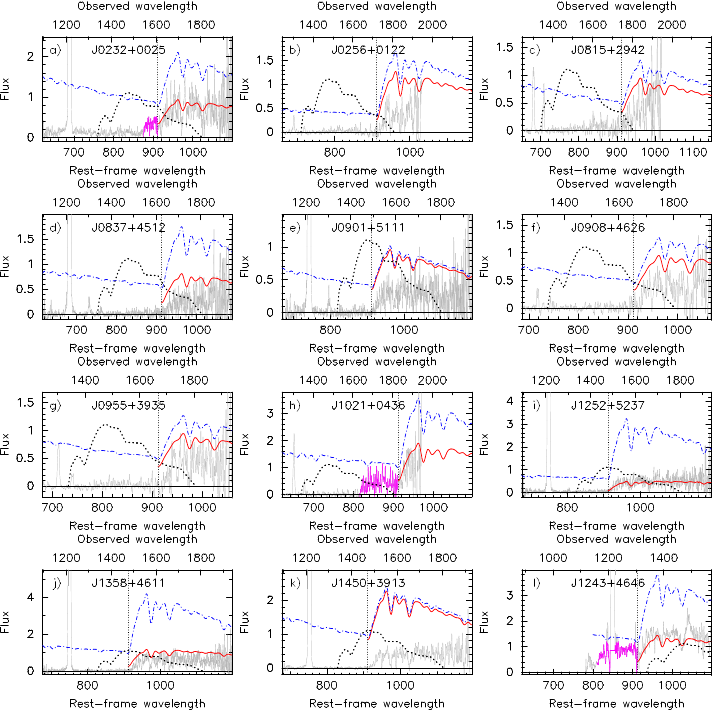}
\caption{Fragments of smoothed spectra in the LyC region for 11 galaxies at
$z$ $\sim$ 0.6 -- 1.0 (a) -- (k) (this paper), and for J1243$+$4646
{(\bf l)} from \citet{I18b}.
Modelled intrinsic SEDs, attenuated by Milky Way extinction 
with $R(V)_\textrm{MW}$ = 3.1 and
internal extinction with $R(V)_\textrm{int}$ = 2.7, assuming the \citet{C89}
reddening law, and the unreddened SEDs are shown
by solid red and dash-dotted blue lines, respectively. Emission features at
the observed wavelengths $\lambda$1216\AA\ and $\lambda$1301\AA\ are geocoronal
Ly$\alpha$ and O~\textsc{i} emission lines, respectively. The parts
of the observed spectra containing the LyC bumps for a) J0232+0025,
h) J1021+0436, and l) J1243+4646 are highlighted by magenta lines.
Dotted black lines in all panels indicate the \textit{GALEX} FUV response curve
reduced to the galaxy rest-frame. The vertical dotted line in all panels
indicates the location of the LyC limit at 912\AA. Fluxes are in 
10$^{-16}$ erg s$^{-1}$ cm$^{-2}$\AA$^{-1}$, and wavelengths are in angstrom. \label{fig4}}
\end{figure*}

The expanded version of the J0232$+$0025 COS spectrum is shown in
Fig.~\ref{fig2}d. 
For comparison, we show the extrapolation of the attenuated SED derived from fitting
the Xshooter spectrum (red line), which is in good agreement with
the observed COS spectrum, and the unattenuated, intrinsic SED (blue line).
The short horizontal red line at wavelengths ranging from 850 to 870\AA\
indicates the flux of the stellar LyC, which is used below to derive
the LyC escape fraction.

\section{Escaping Lyman continuum radiation}\label{sec:lyc}

\citet{I16a,I16b,I18a,I18b,I21a,I22} used the ratio of the escaping fluxes $I_\textrm{esc}$ to the intrinsic fluxes $I_\textrm{mod}$ of the LyC to derive $f_\textrm{esc}$(LyC):
\begin{equation}
f_\textrm{esc}(\textrm{LyC}) =\frac{I_\textrm{esc}(\lambda)}{I_\textrm{mod}(\lambda)}, 
\label{eq:fesc}
\end{equation}
where $\lambda$ is the mean wavelength of the range used for averaging of the LyC flux density (see Table~\ref{tab6}).
We used two methods to derive the intrinsic fluxes $I_\textrm{mod}$ and, correspondingly, the LyC escape fractions $f_\textrm{esc}$(LyC) \citep{I16b}: 
1) from the SED fitting, and 2) from the equivalent width of the H$\beta$ emission line combined with its extinction-corrected flux according to the equation, \begin{equation}
\frac{I(\textrm{H}\beta)}{I_\textrm{mod}(\lambda)}\approx 2.99\times \textrm{EW}(\textrm{H}\beta)^{0.228} \textrm{\AA}.
\end{equation}
However, we note that the SEDs extrapolated to the UV range in Fig.~\ref{fig2} lie somewhat above the observed COS spectra, excluding J1021$+$0436 and J1252$+$5237. On the other hand, the SED derived from the Xshooter spectrum of J0232$+$0025 is in good agreement with the COS spectrum (Fig.~\ref{fig2}c). Therefore, we do not consider the $f_\textrm{esc}$(LyC) derived from SED fitting to be reliable, except for that of J0232$+$0025, J1021$+$0436, and J1252$+$5237. For other galaxies, we consider the $f_\textrm{esc}$(LyC) derived using H$\beta$ emission to be somewhat more reliable, although this method
assumes a young instantaneous burst in the galaxy and ignores the contributions from older populations in both the UV and optical ranges.

The absolute escape fractions of LyC photons, $f_\textrm{esc}$(LyC), derived using these two methods and different wavelength ranges in the LyC (see discussion below), are listed in Table~\ref{tab6}. We find values of $f_\textrm{esc}$(LyC) ranging from 2 to 60~\% for eight out of the eleven galaxies. Among these, three show LyC detections above $3 \sigma$ (J0232+0025, J1021+0436, and J1252+5237), while four galaxies (J0256+0122, J0815+2942, J0901+5111, and J1450+3913) have only upper limits.
We consider the escape fractions $f_\textrm{esc}$(LyC) unreliable for four galaxies (J0837+4512, J0908+4636, J0955+3935, and J1358+4611) with LyC detections below $3 \sigma$. We also note that most of the galaxies with non-detected or unreliably detected stellar LyC emission are characterised by irregular morphologies, likely indicating a stellar origin of the extended emission (Fig.~\ref{fig1}). In contrast, the galaxy J0232+0025, which has the highest stellar LyC escape fraction, is compact and characterised by symmetric, extended, likely Ly$\alpha$, emission (Fig.~\ref{fig3}).

\section{Origin of the LyC bumps} \label{nebular}

We note interesting emission features (LyC bumps) in the LyC near
the Lyman limit in the COS spectra of two galaxies, J0232+0025 and J1021+0436
(Figs.~\ref{fig4}a and \ref{fig4}b).
A bump in the wavelength range 880 -- 912\AA\ is observed in the
spectrum of J0232+0025, whereas a much broader bump in the wavelength range 820~--~912\AA\ is observed in the spectrum of J1021+0436 (magenta lines
in Fig.~\ref{fig4}). Both galaxies are very compact and symmetric
in the COS/NUV acquisition
images, unlike the clumpy galaxies from our sample,
and no other sources are present inside the COS spectroscopic
aperture at angular radii greater than $\sim$ 0\farcs1, corresponding
to linear radii of $\ga$ 0.7 -- 0.8 kpc (Figs.~\ref{fig1}, \ref{fig3}).
However, we note that a faint source at a distance of 0\farcs07 from
the brightest source, corresponding to the projected distance of $\sim$ 0.5 kpc,
is present according to the SB distribution of J1021$+$0436
(Fig.~\ref{figa2}). It could either be located in the galaxy or be a
foreground galaxy at $z$ $\sim$ 0.9. Therefore, we cannot exclude the possibility that the
LyC bump in J1021$+$0436 is unrelated to this galaxy. On the other hand,
no sources are seen in the vicinity of J0232$+$0025
(Figs.~\ref{fig3}, \ref{figa2}),
implying that the bump is likely related only to this galaxy.

The sharp decrease in bumps at $\lambda$880\AA\ in J0232$+$0025 and
at $\lambda$820\AA\ in J1021$+$0436 might be caused by the presence of
intergalactic Lyman limit systems (LLS) at corresponding redshifts of
$\sim$ 0.73 and $\sim$ 0.90. If confirmed, the LLS in the direction of J0232$+$0025
is likely translucent at rest-frame wavelengths $<$ 912\AA, because LyC emission is
detected at $\lambda$ $<$ 880\AA, whereas it appears opaque in the direction of
J1021$+$0436, as LyC emission is not detected at $\lambda$ $<$ 820\AA.
However, the absence of Mg~\textsc{ii} $\lambda$2800 LLS absorption features
bluewards of the Mg~\textsc{ii} $\lambda$2800\AA\ emission line in the Xshooter
spectrum of J0232$+$0025 (Fig.~\ref{figa1}) and in the noisier SDSS
spectrum of J1021$+$0436 (Fig.~\ref{figa3}h) may argue either against
the presence of an LLS or in favour of its very low metallicity.

It is unlikely that bumps can be explained by the presence of dust because the
extinction in both galaxies is very small (Table~\ref{taba3}). Likewise, the
extinction in the LyC range smoothly increases with decreasing wavelength,
attains a maximum at 720\AA, then decreases with decreasing wavelength
\citep{Dr03}. Additionally, the LyC spectrum of J0232$+$0025
at $\lambda$880\AA\ is smooth and lacks any features related to the maximum
in dust extinction at 720\AA.

The smearing of non-ionising galaxy continuum also cannot play a role below
900\AA\ because the line spread function (LSF) of G140L has a sharp core
with high values within $\pm$ 1\AA, and extended wings with low values. As a result, it is
not able to redistribute non-ionising emission to wavelengths below 900\AA.

We suggest that the two bumps are due to nebular
LyC emission. They may be connected with the high EW(H$\beta$) $>$ 200\AA\
(Table \ref{taba3}) and thus with a high fraction of nebular continuum
($\ga$ 20\%) near the H$\beta$ emission line. However, we find no evidence of a bump
in the spectrum of J0837$+$4512 (Fig.~\ref{fig4}d) with the highest
EW(H$\beta$) of 368\AA\ (Table \ref{taba3}).
The average monochromatic LyC fluxes in the bumps
of J0232+0025 and J1021+0436 are considerably larger
than that at shorter wavelengths. However, no such emission is detected in the
spectra of the other nine galaxies with similar integrated characteristics
(Table~\ref{tab1}). The only difference is that the stellar LyC escape fraction
in J0232+0025 is considerably higher than in other galaxies (Table~\ref{tab6}),
implying that the neutral gas in this galaxy is translucent to LyC emission. 
However,  in the spectrum of J1021+0436 no flux is detected in the LyC at
wavelengths $\la 820$\AA.

\begin{figure*}
\centering
\includegraphics[angle=0,width=0.75\linewidth]{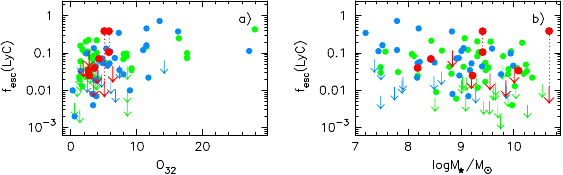}
\caption{Relations between LyC escape fraction
$f_\textrm{esc}$(LyC) in low-redshift LyC leaking galaxies derived from SED
fits and a) the O$_{32}$ =
[O~\textsc{iii}]$\lambda$5007/[O~\textsc{ii}]$\lambda$3727 
emission-line flux ratio, and b) stellar mass $M_\star$. 
LyC leakers from \citet{I16a,I16b,I18a,I18b,I21a}, \citet{B14}, 
and \citet{C17} are shown by blue-filled circles.
The galaxies from \citet{Fl22a,Fl22b} and this paper with detected LyC emission
are represented by green- and red-filled circles, respectively. The blue,
green, and red arrows denote upper limits in galaxies with non-detected
LyC emission.
Dotted red lines connect the two different values
of the LyC escape fraction from Table~\ref{tab6}
for J0232+0025 and J1021+0436, corresponding to higher and lower values
obtained from the fluxes within the wavelength ranges with and without the LyC bump,
respectively.
\label{fig5}}
\end{figure*}

The properties of nebular LyC bump were first considered by \citet{In10}, who
was motivated by the enhanced level of LyC emission just bluewards of
the Lyman limit at
$\lambda$=912\AA, observed in a number of SFGs at $z$ $\sim$ 3. He showed
that the escaping hydrogen bound-free emission produces a strong bump just below
the Lyman limit
due to the radiation energy redistribution of stellar LyC by nebulae. The
effect depends on the optical depth (or column density) of neutral hydrogen
and is maximal at intermediate column densities. The bump is absent at very low neutral hydrogen column densities when the LyC escape fraction is high
($f_\textrm{esc}$(LyC)~$\sim$~1) and at high neutral
hydrogen column densities when $f_\textrm{esc}$(LyC)~$\sim$~0.

\citet{Si24} further considered the properties of the nebular LyC bump and
found that it is strongest at $N$(H~\textsc{i}) $\sim$ 10$^{17}$ --
10$^{18}$ cm$^{-2}$, where escape fractions, calculated as the ratio between the
intrinsic stellar spectrum and the emergent LyC stellar plus nebular spectrum
over the same wavelength range, are increased by more than a factor of 3, compared to escape fractions of the transmitted stellar spectrum.
This implies that some of LyC photons detected in the literature may have a nebular, rather than stellar, origin.

The presence of the nebular
LyC emission implies that LyC escape fractions in the LyC leaking galaxies at
lower redshifts of 0.3 -- 0.4 might be somewhat overestimated.
This is because the LyC flux in galaxies at $z$ $\sim$ 0.3 -- 0.4 can only be
observed and measured in \textit{HST}/COS spectra at rest-frame wavelengths
$\lambda$~$\ga$~870~--~900\AA\ near the Lyman limit, where contributions from
nebular LyC emission could be important but are not distinguished from stellar emission. Additionally, contamination of LyC emission
by geocoronal Ly$\alpha$ emission in galaxies with $z$ $\sim$ 0.32 -- 0.36
is high. Therefore, the fraction of the LyC leaking galaxies
at $z$ $<$ 0.4 with LyC nebular emission is difficult to estimate. However, at
higher redshifts, ionising radiation at shorter rest-frame wavelengths can be
observed with the \textit{HST}/COS. 
To illustrate this, we add in Fig.~\ref{fig4}l the spectrum of J1243+4646 at
$z=0.4317$, which shows one of the highest LyC escape fractions
$f_\textrm{esc}$(LyC)~$\sim$~0.7 derived by \citet{I18b} near the Lyman limit. 
This galaxy has the highest redshift $z$ among the galaxies observed by
\citet{I18b}; therefore, the \textit{HST}/COS spectrum extends to shorter
wavelengths compared with other galaxies from their sample.
Also, the equivalent width of the H$\beta$ emission line in the spectrum
of J1243+4646 is above 200\AA, similar to that in the spectra of the two
galaxies with the LyC bumps discussed in this paper, and there are no other sources in
the vicinity of this galaxy \citep[Figs.~4c and 5c in ][]{I18b}.
As seen in Fig.~\ref{fig4}l, the LyC continuum is high and nearly
constant at $\lambda$ $>$ 820\AA, then decreases rapidly
at shorter wavelengths, similar to the behaviour observed in J0232+0025
(Fig.~\ref{fig4}a) and J1021+0436
(Fig.~\ref{fig4}h). We also note that the energy distribution of the nebular
LyC in all three galaxies in Fig.~\ref{fig4} is nearly constant,
whereas the modelled energy distribution \citep{In10,Si24} has a sharp peak
at 912\AA\ and steep decrease at shorter wavelengths. This discrepancy
between models and observations warrants detailed investigation in a separate
study, which is beyond the scope of this paper.
Finally, we note that the total LyC escape fraction predicted using the 
observed flux ratio $F_{700\AA}/F_{1100\AA}$ and the expression proposed
by \cite{Si24} from their CLOUDY models, assuming isotropic escape,
yields $f_\textrm{esc}$(LyC)$\approx$1
for J0232+0025. This is likely incompatible with the presence of strong
emission lines in this galaxy as well as with LyC nebular free-bound emission.

The effect of nebular LyC can likely be seen in distributions between the
observed LyC escape fractions and the LyC escape fractions predicted by
\citet{J24}, who use combinations of several global characteristics of low-redshift
($z$ $\sim$ 0.3 -- 0.4) LyC emitters. \citet{J24} developed and analysed new
multivariate predictors of $f_\textrm{esc}$(LyC) based on the Cox
proportional hazards model, a survival analysis technique that incorporates
both detections and upper limits. They found that in strong LyC leakers with
$f_\textrm{esc}$(LyC) $\ga$ 0.2, the predicted escape fractions are systematically
lower than the observed ones. \citet{J24} suggest
that this difference is caused by unaccounted parameter(s),
such as the H\,\textsc{i} covering fraction and anisotropy of LyC escape
\citep[see also earlier studies by ][]{Z13,Z17,Fl22b,Sal22}.
We also suggest that such a difference is due to the
contribution of nebular LyC
components to the observed $f_\textrm{esc}$(LyC) near the Lyman limit, which is not accounted for in the \citet{J24} analysis.

If confirmed, the presence of nebular LyC emission in LyC galaxies is expected to be rare,
as its emergence requires special conditions \citep[e.g. ][]{Si24}.
In particular, stacks of a large number of LzLCS+ galaxies in the wavelength
range 820 -- 900\AA\ show no evidence of nebular contribution
\citep{Fl25}. The higher EW(H$\beta$) $>$ 200\AA\ in all three galaxies
with suspected nebular LyC emission, combined with their very compact structure compared to the average LzLCS+ galaxy, may be necessary but apparently not sufficient conditions.

\section{Indicators of high LyC escape fraction} \label{Ind}

Since direct measurements of LyC emission in high-$z$ galaxies are basically
unfeasible but of great interest, 
several indirect methods to determine LyC emission from SFGs
at low-redshift have been proposed. These include \citet{JO13,NO14,F16,I16a,I16b,I18a,I18b,I21a,Fl22a,Fl22b,J24},
who determined relations between the LyC escape fraction
$f_\textrm{esc}$(LyC) and such various observable or derivable characteristics from the spectra in the UV and optical ranges as stellar mass
$M_\star$, Ly$\alpha$ profile, Ly$\alpha$ escape fraction, and O$_{32}$ ratio.
However, at redshifts greater than 0.6, observing the Ly$\alpha$ profile -- the best indicator --
is not possible with the \textit{HST} COS/FUV. On the other hand,
such characteristics as $M_\star$ and O$_{32}$ obtained from the optical data are
available.

In addition to the 11 new LyC measurements presented above, we have
a sample of approximately one hundred low-$z$ galaxies with a wide range of
O$_{32}$, up to 29, and directly derived $f_\textrm{esc}$(LyC)
\citep{B14,L16,I16a,I16b,C17,I18a,I18b,I21a,I22,Fl22a,Fl22b}. 
The relation between 
$f_\textrm{esc}$(LyC) and O$_{32}$ is presented in Fig.~\ref{fig5}a. 
There is a trend of increasing $f_\textrm{esc}$(LyC) with increasing 
O$_{32}$, albeit with a substantial scatter.
This large scatter is due to the dependence of O$_{32}$ on other parameters such
as metallicity, hardness of ionising radiation, and ionisation
\citep[e.g. ][]{I21a,I22,Fl22b,Red23}.
Additionally, the spread of 
$f_\textrm{esc}$(LyC) can also be due to varying degrees of leakage through
channels with low optical depth and to different orientation of these channels
relative to the observer's line of sight. 
Therefore, a high O$_{32}$ is not a very certain indicator of high 
$f_\textrm{esc}$(LyC), and it is only a necessary condition for escaping 
radiation \citep[e.g. ][]{I18b,Na20}. The only 
firm result from Fig.~\ref{fig5}a is 
that $f_\textrm{esc}$(LyC) is low ($<$ 10\%) on average, and that the fraction of
LyC non-detections is higher in objects with O$_{32}$~$<$~10
\citep[see also][]{Fl22a,Fl22b}.

It has also been suggested that $f_\textrm{esc}$(LyC) tends to be higher in 
low-mass galaxies \citep{W14,T17}. 
Previous observations have shown a slight tendency for $f_\textrm{esc}$(LyC) to 
increase with decreasing stellar mass, albeit with large intrinsic scatter. 
We present in Fig.~\ref{fig5}b the relation between $f_\textrm{esc}$(LyC) and
stellar mass $M_\star$, including our new data. The additional data shows weak dependence of $f_\textrm{esc}$(LyC) on $M_\star$.

\section{Conclusions}\label{summary}

We presented new \textit{Hubble Space Telescope} (\textit{HST}) Cosmic Origins
Spectrograph (COS) observations of eleven low-mass SFGs
at redshifts, $z$, in the range 0.6145 -- 1.0053. These redshifts are approximately a
factor of 2 higher than those of the LyC leaking galaxies studied so far with
the \textit{HST}/COS \citep[e.g.][]{I16a,I16b,I18a,I18b,I21a,I22,Fl22a,Fl22b}, extending detections of LyC emission with COS to $z=1$, corresponding to $\sim 5.9$ Gyr after the Big Bang.
Consequently, the observed wavelength range for the LyC in galaxies from this
paper is considerably larger, starting at rest-frame wavelengths of
$\sim$~600~--~700\AA\ , compared to $\sim$~800~--~850\AA\ in previous studies.
This reveals, for the first time, a great diversity in the spectral shapes of the
LyC in galaxies with a considerable escape of ionising radiation.
In particular, the nebular contribution to LyC emission near the Lyman limit,
predicted by earlier studies \citep{In10,Si24}, can now be confirmed
and studied for the first time.
We supplemented the \textit{HST} data with the SDSS spectra for all galaxies and
the VLT/Xshooter spectrum for J0232$+$0025 to produce SEDs spanning the entire UV and optical ranges and to derive the
intrinsic flux in the LyC.
Our main results are summarised as follows:

1. Emission of the LyC is detected in seven out of the eleven
galaxies with the escape fraction $f_\textrm{esc}$(LyC) in the range between approximately
2 and 60~\%. Upper limits of $f_\textrm{esc}$(LyC) were obtained for the remaining
galaxies.

2. A complex bump-shaped structure is found in the LyC spectrum in the rest-frame
wavelength range 870~--~912\AA\ of the spectrum of J0232$+$0025 and in the rest-frame wavelength
range 820~--~912\AA\ of the spectrum of J1021$+$0436. 
We assume this emission to be nebular, with the monochromatic
flux density of the nebular continuum approximately twice that of the stellar LyC emission in J0232$+$0025, 
and much higher in J1021$+$0436.
The possible presence of nebular emission may pose a problem in
determining the LyC escape fraction in galaxies at
$z$~$\sim$~0.3--0.4, since only a narrow part of the LyC is generally observed
at rest-frame wavelengths ranging from $\sim$~850~to~912\AA.
These two varying components (stellar and nebular LyC) can contribute to
the scatter in $f_\textrm{esc}$(LyC), which was not measured homogeneously due to
observational limitations.

This study demonstrates the importance of studying the spectral shape of the
LyC spectrum over a broad wavelength range, in order to unveil the origin of
ionising photons escaping from galaxies, and ultimately, refine our
understanding of the nature of the sources of reionisation.

\begin{acknowledgements}
Based on observations made with the NASA/ESA \textit{Hubble Space Telescope}, 
obtained from the data archive at the Space Telescope Science Institute. 
STScI is operated by the Association of Universities for Research in Astronomy,
Inc. under NASA contract NAS 5-26555. Support for this work was provided by 
NASA through grant numbers HST-GO-16271, HST-GO-17169, and HST-GO-17171
from the Space
Telescope Science Institute, which is operated by AURA, Inc., under NASA
contract NAS 5-26555. YI, DS, and NG acknowledge support from the joint
Ukrainian-Swiss project No. 224866. YI and NG acknowledge support from the
Simons Foundation.
CS acknowledges support from the Science and Technology Facilities Council
(STFC), by the ERC through Advanced Grant 695671 “QUENCH”, by the UKRI Frontier
Research grant RISEandFALL.
Funding for SDSS-III has been provided by the Alfred P. Sloan Foundation, 
the Participating Institutions, the National Science Foundation, and the U.S. 
Department of Energy Office of Science. The SDSS-III web site is 
http://www.sdss3.org/. SDSS-III is managed by the Astrophysical Research 
Consortium for the Participating Institutions of the SDSS-III Collaboration. 
\textsc{iraf} is distributed by 
the National Optical Astronomy Observatories, which are operated by the 
Association of Universities for Research in Astronomy, Inc., under cooperative 
agreement with the National Science Foundation. \textsc{stsdas} is a product of 
the Space Telescope Science Institute, which is operated by AURA for NASA.
GALEX is a NASA mission  managed  by  the  Jet  Propulsion  Laboratory.
This research has made use of the NASA/IPAC Extragalactic Database (NED) which 
is operated by the Jet  Propulsion  Laboratory,  California  Institute  of  
Technology,  under  contract with the National Aeronautics and Space 
Administration.
\end{acknowledgements}

\bibliographystyle{aa}

\begin{appendix}
\onecolumn

\section{Tables and figures}

  \begin{table*}
  \caption{Apparent AB magnitudes with errors in parentheses, compiled from SDSS and \textit{GALEX} databases
\label{taba1}}
\centering
\begin{tabular}{lcccccccc} \hline
Name&\multicolumn{5}{c}{SDSS}
&&\multicolumn{2}{c}\textit{GALEX}\\
    &\multicolumn{1}{c}{$u$}&\multicolumn{1}{c}{$g$}&\multicolumn{1}{c}{$r$}&\multicolumn{1}{c}{$i$}&\multicolumn{1}{c}{$z$}&&FUV&NUV
\\
    &\multicolumn{1}{c}{(err)}&\multicolumn{1}{c}{(err)}&\multicolumn{1}{c}{(err)}&\multicolumn{1}{c}{(err)}&\multicolumn{1}{c}{(err)}&&(err)&(err) \\
\hline
J0232+0025& 21.31& 21.34& 21.18& 21.38& 20.18&& 22.40& 20.94\\
            &(0.11)&(0.04)&(0.05)&(0.09)&(0.14)&&(0.14)&(0.05)\\
J0256+0122& 21.11& 20.98& 20.90& 20.61& 20.54&&  ... & 21.57\\
            &(0.08)&(0.03)&(0.04)&(0.04)&(0.11)&& (...)&(0.07)\\
J0815+2942& 21.74& 21.33& 21.25& 21.00& 20.71&&  ... & 21.54\\
            &(0.15)&(0.04)&(0.06)&(0.06)&(0.19)&& (...)&(0.21)\\
J0837+4512& 21.82& 21.63& 21.43& 21.26& 20.13&& 24.24& 21.75\\
            &(0.14)&(0.07)&(0.07)&(0.08)&(0.11)&&(0.35)&(0.07)\\
J0901+5111& 21.85& 21.66& 21.37& 20.91& 20.99&& 22.15& 22.16\\
            &(0.16)&(0.06)&(0.07)&(0.08)&(0.32)&&(0.53)&(0.35)\\
J0908+4626& 21.62& 21.41& 21.31& 21.07& 20.65&&  ... & 21.61\\
            &(0.17)&(0.06)&(0.08)&(0.10)&(0.30)&& (...)&(0.11)\\
J0955+3935& 21.31& 21.41& 21.44& 21.16& 20.90&&  ... & 21.11\\
            &(0.11)&(0.06)&(0.09)&(0.10)&(0.26)&& (...)&(0.31)\\
J1021+0436& 20.80& 20.63& 20.72& 20.56& 20.42&&  ... & 20.75\\
            &(0.07)&(0.02)&(0.03)&(0.04)&(0.13)&& (...)&(0.05)\\
J1252+5237& 21.64& 21.87& 21.64& 21.04& 21.04&&  ... & 21.77\\
            &(0.13)&(0.06)&(0.07)&(0.06)&(0.22)&& (...)&(0.14)\\
J1358+4611& 21.42& 21.44& 21.16& 20.67& 21.00&& 22.90& 21.29\\
            &(0.09)&(0.04)&(0.05)&(0.05)&(0.24)&&(0.24)&(0.09)\\
J1450+3913& 21.33& 21.38& 21.13& 20.47& 20.64&&  ... & 21.64\\
            &(0.10)&(0.05)&(0.05)&(0.05)&(0.16)&& (...)&(0.22)\\
\hline
\end{tabular}

  \end{table*}

\begin{figure}
\centering
\includegraphics[angle=0,width=0.99\linewidth]{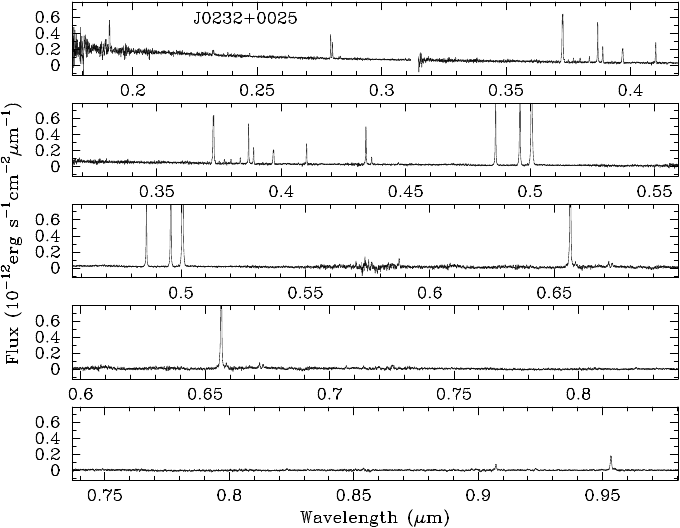}
\caption{Rest-frame Xshooter spectrum of J0232+0025. Fluxes are in 
10$^{-12}$ erg s$^{-1}$ cm$^{-2}$$\mu$m$^{-1}$, wavelengths are in micrometers.
\label{figa1}}
\end{figure}

\begin{figure*}
\centering
\includegraphics[angle=0,width=0.99\linewidth]{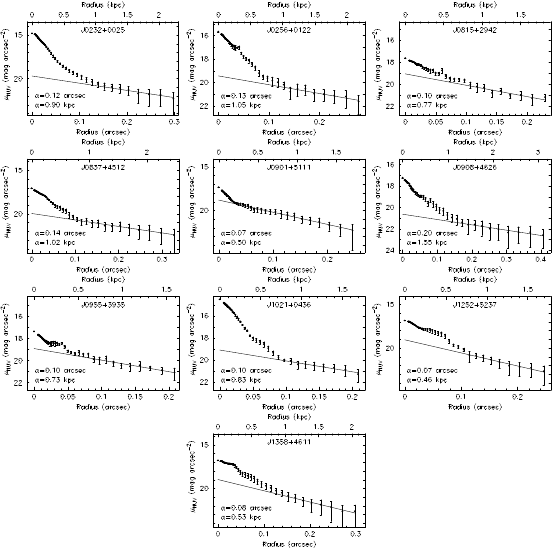}
\caption{NUV SB profiles of galaxies (black-filled circles with 1$\sigma$ error bars). Solid lines are linear fits to the outer regions of the galaxies.
\label{figa2}}
\end{figure*}

\begin{table*}
\centering
\caption{Extinction-corrected emission-line flux ratios and rest-frame equivalent widths in the Xshooter spectrum of J0232$+$0025 \label{taba2}}
\begin{tabular}{lrrclrr} \hline
Line& \multicolumn{1}{c}{100$\times$$I$($\lambda$)/$I$(H$\beta$)} & \multicolumn{1}{c}{EW($\lambda$)$^\textrm{a}$}&&Line& \multicolumn{1}{c}{100$\times$$I$($\lambda$)/$I$(H$\beta$)} & \multicolumn{1}{c}{EW($\lambda$)$^\textrm{a}$}  \\ \hline
1906.68 C \textsc{iii}]            & 35.05$\pm$3.36&  4.39$\pm$0.22&&4363.21 [O \textsc{iii}]           &  9.99$\pm$0.39& 10.14$\pm$0.15\\
1908.73 C \textsc{iii}]            & 21.62$\pm$2.60&  2.64$\pm$0.24&&4471.48 He \textsc{i}              &  3.49$\pm$0.21&  3.90$\pm$0.21\\
2382.00 Fe \textsc{ii}             &  4.94$\pm$0.61&  0.91$\pm$0.10&&4685.94 He \textsc{ii}             &  1.19$\pm$0.14&  1.25$\pm$0.17\\
2626.00 Fe \textsc{ii}             &  2.43$\pm$0.31&  0.58$\pm$0.07&&4861.33 H$\beta$                &100.00$\pm$2.85&217.40$\pm$0.31\\
2795.50 Mg \textsc{ii}             & 28.40$\pm$1.25&  8.16$\pm$0.09&&4958.92 [O \textsc{iii}]           &233.05$\pm$6.64&350.70$\pm$1.64\\
2802.70 Mg \textsc{ii}             & 17.09$\pm$0.76&  4.85$\pm$0.07&&5006.80 [O \textsc{iii}]           &708.32$\pm$20.1&951.50$\pm$0.55\\
3188.00 He \textsc{i}              &  2.75$\pm$0.39&  1.07$\pm$0.12&&5754.64 [N \textsc{ii}]            &  8.41$\pm$3.53&  9.24$\pm$0.12\\
3727.00 [O \textsc{ii}]            &124.54$\pm$4.07& 90.90$\pm$0.72&&5875.60 He \textsc{i}              & 14.25$\pm$1.94& 28.69$\pm$0.17\\
3750.15 H12                     &  3.45$\pm$0.44&  2.40$\pm$0.18&&6548.10 [N~\textsc{ii}]            &  7.90$\pm$0.82& 30.03$\pm$2.03\\
3770.63 H11                     &  5.56$\pm$0.51&  4.07$\pm$0.25&&6562.80 H$\alpha$               &283.18$\pm$9.29&963.40$\pm$2.08\\
3797.90 H10                     &  6.04$\pm$0.43&  4.16$\pm$0.20&&6583.40 [N~\textsc{ii}]            & 17.34$\pm$1.14& 60.78$\pm$2.18\\
3835.39 H9                      &  7.63$\pm$0.38&  5.26$\pm$0.14&&6716.40 [S~\textsc{ii}]            &  9.74$\pm$0.57& 19.08$\pm$1.09\\
3868.76 [Ne \textsc{iii}]          & 52.16$\pm$1.64& 38.99$\pm$0.13&&6730.80 [S~\textsc{ii}]            &  7.92$\pm$0.57& 15.20$\pm$1.01\\
3889.00 He \textsc{i}+H8           & 23.25$\pm$0.78& 20.73$\pm$0.16&&7065.30 He \textsc{i}              &  4.81$\pm$0.44& 16.04$\pm$1.01\\
3968.00 [Ne \textsc{iii}]+H7       & 36.15$\pm$1.18& 30.86$\pm$0.31&&7135.80 [Ar~\textsc{iii}]          &  5.16$\pm$0.39& 15.11$\pm$1.50\\
4026.19 He \textsc{i}              &  2.20$\pm$0.26&  1.85$\pm$0.23&&9069.00 [S~\textsc{iii}]           & 16.15$\pm$0.70&136.00$\pm$2.52\\
4101.74 H$\delta$               & 27.21$\pm$0.88& 24.30$\pm$0.22&&9530.60 [S~\textsc{iii}]           & 40.62$\pm$1.63&794.90$\pm$4.34\\
4340.47 H$\gamma$               & 49.76$\pm$1.48& 49.48$\pm$0.20\\ \\

&&\multicolumn{1}{l}{$C$(H$\beta$)$^\textrm{b}$}         &&\multicolumn{1}{c}{0.085$\pm$0.039}\\
&&\multicolumn{1}{l}{$F$(H$\beta$)$^\textrm{c}$}        &&\multicolumn{1}{c}{7.40$\pm$0.14}\\
&&\multicolumn{1}{l}{EW(abs)$^\textrm{d}$}                &&\multicolumn{1}{c}{2.00$\pm$3.38}\\
\hline
\end{tabular}
\tablefoot{$^\textrm{a}$In \AA. $^\textrm{b}$Extinction coefficient, derived from the observed hydrogen Balmer decrement. $^\textrm{c}$Observed flux in units of 10$^{-16}$ erg s$^{-1}$ cm$^{-2}$.}\\ \hbox{$^\textrm{d}$Equivalent width of hydrogen absorption lines in \AA.}
\end{table*}

\begin{table*}
\centering
\caption{Extinction-corrected fluxes and restframe equivalent widths of the emission lines in SDSS spectra
\label{taba3}}
\begin{tabular}{lcrrrrrrrr} \hline
 & &\multicolumn{8}{c}{Galaxy}\\
&&\multicolumn{2}{c}{J0232$+$0025}&\multicolumn{2}{c}{J0256$+$0122}&\multicolumn{2}{c}{J0815$+$2942}& \multicolumn{2}{c}{J0837$+$4512}\\
Line &\multicolumn{1}{c}{$\lambda$}& \multicolumn{1}{c}{$I$$^\textrm{a}$}&\multicolumn{1}{c}{EW$^\textrm{b}$}&\multicolumn{1}{c}{$I$$^\textrm{a}$}&\multicolumn{1}{c}{EW$^\textrm{b}$}&\multicolumn{1}{c}{$I$$^\textrm{a}$}&\multicolumn{1}{c}{EW$^\textrm{b}$}&\multicolumn{1}{c}{$I$$^\textrm{a}$}&\multicolumn{1}{c}{EW$^\textrm{b}$}\\
\hline
Mg~\textsc{ii}          &2796&  29.6$\pm$6.0&  14& 25.5$\pm$5.0&   6& 39.2$\pm$15.&  22& 12.2$\pm$7.1&  10\\
Mg~\textsc{ii}          &2803&  18.1$\pm$5.5&   8& 21.9$\pm$4.8&   5& 20.6$\pm$14.&  10&  8.2$\pm$7.0&   6\\
$[$O~\textsc{ii}$]$     &3727& 104.7$\pm$8.1& 130&169.3$\pm$11.&  85&248.9$\pm$24.& 232&144.7$\pm$10.& 258\\
H9                   &3836&   7.7$\pm$4.2&  13&   ...       & ...&   ...       & ...&  8.0$\pm$6.0&  46\\
$[$Ne~\textsc{iii}$]$   &3869&  46.4$\pm$5.7&  68& 43.2$\pm$5.1&  38& 24.1$\pm$12.&  21& 45.8$\pm$6.7& 104\\
H8+He~\textsc{i}        &3889&  20.0$\pm$4.7&  31& 20.0$\pm$3.9&  16& 25.1$\pm$12.&  22& 18.1$\pm$6.4&  48\\
H7+$[$Ne~\textsc{iii}$]$&3969&  30.5$\pm$5.0&  50& 29.6$\pm$4.4&  26& 24.2$\pm$12.&  27& 30.8$\pm$6.7&  52\\
H$\delta$            &4101&  31.2$\pm$5.0&  53& 31.9$\pm$4.2&  30& 26.2$\pm$12.&  24& 25.1$\pm$6.2&  58\\
H$\gamma$            &4340&  48.2$\pm$5.5&  95& 48.6$\pm$5.2&  47& 49.1$\pm$12.&  56& 47.6$\pm$6.4& 141\\
$[$O~\textsc{iii}$]$    &4363&   7.1$\pm$3.8&  14&  7.0$\pm$3.1&   5&  3.8$\pm$1.0&   5&  8.7$\pm$4.9&  23\\
H$\beta$             &4861& 100.0$\pm$6.9& 281&100.0$\pm$7.1& 156&100.0$\pm$13.& 193&100.0$\pm$7.3& 368\\
$[$O~\textsc{iii}$]$    &4959& 200.7$\pm$9.8& 610&156.9$\pm$9.1& 250&159.6$\pm$15.& 157&177.9$\pm$9.2& 634\\
$[$O~\textsc{iii}$]$    &5007& 609.9$\pm$20.&1329&452.0$\pm$17.& 450&502.6$\pm$29.& 548&514.3$\pm$18.& 967\\
$C$(H$\beta$)$^\textrm{c}$  &&\multicolumn{2}{c}{0.000$\pm$0.090}&\multicolumn{2}{c}{0.000$\pm$0.092}&\multicolumn{2}{c}{0.000$\pm$0.167}&\multicolumn{2}{c}{0.084$\pm$0.093}\\
EW(abs)$^\textrm{d}$      &&\multicolumn{2}{c}{0.0}&\multicolumn{2}{c}{0.0}&\multicolumn{2}{c}{0.0}&\multicolumn{2}{c}{1.0} \\
$F$(H$\beta$)$^\textrm{e}$     &&\multicolumn{2}{c}{7.4$\pm$0.3}&\multicolumn{2}{c}{6.0$\pm$0.6}&\multicolumn{2}{c}{3.5$\pm$0.2}&\multicolumn{2}{c}{8.1$\pm$0.3}\\
\hline
&&\multicolumn{2}{c}{J0901+5111}&\multicolumn{2}{c}{J0908$+$4626}&\multicolumn{2}{c}{J0955$+$3935}& \multicolumn{2}{c}{J1021$+$0436}\\
Line &\multicolumn{1}{c}{$\lambda$}&\multicolumn{1}{c}{$I$$^\textrm{a}$}&\multicolumn{1}{c}{EW$^\textrm{b}$}&\multicolumn{1}{c}{$I$$^\textrm{a}$}&\multicolumn{1}{c}{EW$^\textrm{b}$}&\multicolumn{1}{c}{$I$$^\textrm{a}$}&\multicolumn{1}{c}{EW$^\textrm{b}$}&\multicolumn{1}{c}{$I$$^\textrm{a}$}&\multicolumn{1}{c}{EW$^\textrm{b}$}\\
\hline
Mg~\textsc{ii}          &2796&  18.2$\pm$4.7&  25& 19.6$\pm$3.8&  14& 24.7$\pm$4.5&  14& 18.2$\pm$6.6&   9\\
Mg~\textsc{ii}          &2803&  11.9$\pm$4.4&  14& 14.7$\pm$3.5&   8& 17.9$\pm$4.0&  10& 14.9$\pm$6.5&   8\\
$[$O~\textsc{ii}$]$     &3727& 113.4$\pm$8.1& 253&111.7$\pm$7.9& 103&176.5$\pm$11.& 120&126.2$\pm$8.6& 160\\
H9                   &3836&   6.2$\pm$3.5&  14&   ...       & ...&   ...       & ...&   ...       & ...\\
$[$Ne~\textsc{iii}$]$   &3869&  45.7$\pm$5.3&  79& 38.0$\pm$4.0&  47& 35.3$\pm$4.8&  31& 41.0$\pm$6.7&   62\\
H8+He~\textsc{i}        &3889&  19.4$\pm$4.1&  64& 19.7$\pm$12.&  17& 17.9$\pm$12.&   9& 15.8$\pm$5.2&  24\\
H7+$[$Ne~\textsc{iii}$]$&3969&  32.9$\pm$4.7&  80& 25.7$\pm$16.&  25& 24.6$\pm$11.&  27& 26.9$\pm$5.4&  41\\
H$\delta$            &4101&  27.2$\pm$4.4&  73& 24.5$\pm$12.&  23& 24.2$\pm$9.1&  20& 30.5$\pm$5.3&  53\\
H$\gamma$            &4340&  48.7$\pm$5.1& 142& 34.6$\pm$14.&  26& 41.8$\pm$8.5&  42& 45.6$\pm$5.5&  77\\
$[$O~\textsc{iii}$]$    &4363&  13.4$\pm$3.5&  59&  3.9$\pm$2.5&   3&  9.6$\pm$3.5&   6&  4.3$\pm$3.0&   9\\
H$\beta$             &4861& 100.0$\pm$6.8& 281&100.0$\pm$12.& 112&100.0$\pm$8.6& 160&100.0$\pm$6.5& 206\\
$[$O~\textsc{iii}$]$    &4959& 250.8$\pm$11.& 730&142.3$\pm$8.0& 208&161.3$\pm$10.& 217&204.9$\pm$9.0& 350\\
$[$O~\textsc{iii}$]$    &5007& 722.4$\pm$16.& 891&425.7$\pm$16.&1045&464.7$\pm$18.& 435&636.3$\pm$19.& 907\\
$C$(H$\beta$)$^\textrm{c}$  &&\multicolumn{2}{c}{0.000$\pm$0.088}&\multicolumn{2}{c}{0.050$\pm$0.084}&\multicolumn{2}{c}{0.000$\pm$0.093}&\multicolumn{2}{c}{0.065$\pm$0.084}\\
EW(abs)$^\textrm{d}$      &&\multicolumn{2}{c}{0.0}&\multicolumn{2}{c}{3.0}&\multicolumn{2}{c}{3.6}&\multicolumn{2}{c}{0.5} \\
$F$(H$\beta$)$^\textrm{e}$     &&\multicolumn{2}{c}{7.4$\pm$0.3}&\multicolumn{2}{c}{7.4$\pm$0.6}&\multicolumn{2}{c}{6.0$\pm$0.6}&\multicolumn{2}{c}{10.5$\pm$0.3}\\
\hline
&&\multicolumn{2}{c}{J1252$+$5237}& \multicolumn{2}{c}{J1358$+$4611}& \multicolumn{2}{c}{J1450$+$3913}\\
Line &\multicolumn{1}{c}{$\lambda$}& \multicolumn{1}{c}{$I$$^\textrm{a}$}&\multicolumn{1}{c}{EW$^\textrm{b}$}&\multicolumn{1}{c}{$I$$^\textrm{a}$}&\multicolumn{1}{c}{EW$^\textrm{b}$}&\multicolumn{1}{c}{$I$$^\textrm{a}$}&\multicolumn{1}{c}{EW$^\textrm{b}$}\\
\cline{1-8}
Mg~\textsc{ii}          &2796&  39.9$\pm$6.8&  14& 29.7$\pm$4.2&  10& 10.0$\pm$2.2&  13\\
Mg~\textsc{ii}          &2803&  30.4$\pm$6.2&  11& 18.7$\pm$3.6&   6&  9.4$\pm$2.3&  15\\
$[$O~\textsc{ii}$]$     &3727& 133.8$\pm$9.8&  95&165.6$\pm$8.9& 100&133.7$\pm$7.5& 106\\
H9                   &3836&   8.7$\pm$3.5&   8& 12.2$\pm$12.&   7&  9.7$\pm$6.1&   7\\
$[$Ne~\textsc{iii}$]$   &3869&  45.4$\pm$5.7&  32& 36.3$\pm$3.9&  28& 29.5$\pm$3.5&  24\\
H8+He~\textsc{i}        &3889&  14.4$\pm$4.0&  11& 19.6$\pm$15.&  11& 17.3$\pm$7.4&  36\\
H7+$[$Ne~\textsc{iii}$]$&3969&  27.4$\pm$4.7&  23& 31.3$\pm$13.&  27& 27.0$\pm$6.9&  21\\
H$\delta$            &4101&  27.4$\pm$4.7&  26& 31.4$\pm$11.&  31& 25.2$\pm$6.5&  11\\
H$\gamma$            &4340&  47.4$\pm$5.2&  67& 45.7$\pm$13.&  40& 47.2$\pm$7.1&  41\\
$[$O~\textsc{iii}$]$    &4363&   8.9$\pm$3.1&  11&  5.9$\pm$2.1&   6&  5.7$\pm$2.0&   7\\
H$\beta$             &4861& 100.0$\pm$6.9& 137&100.0$\pm$10.& 154&100.0$\pm$6.9& 135\\
$[$O~\textsc{iii}$]$    &4959& 178.9$\pm$9.4& 203&163.8$\pm$7.4& 215&157.8$\pm$7.4& 231\\
$[$O~\textsc{iii}$]$    &5007& 571.2$\pm$20.&1151&469.9$\pm$13.& 475&463.7$\pm$15.& 548\\
$C$(H$\beta$)$^\textrm{c}$  &&\multicolumn{2}{c}{0.236$\pm$0.090}&\multicolumn{2}{c}{0.136$\pm$0.071}&\multicolumn{2}{c}{0.000$\pm$0.073}\\
EW(abs)$^\textrm{d}$      &&\multicolumn{2}{c}{0.0}&\multicolumn{2}{c}{3.2}&\multicolumn{2}{c}{2.6} \\
$F$(H$\beta$)$^\textrm{e}$     &&\multicolumn{2}{c}{6.5$\pm$0.6}&\multicolumn{2}{c}{10.4$\pm$0.7}&\multicolumn{2}{c}{9.8$\pm$0.7}\\
\cline{1-8}
\end{tabular}
\tablefoot{$^\textrm{a}$$I$=100$\times$$I$($\lambda$)/$I$(H$\beta$): $I$($\lambda$) and $I$(H$\beta$) are extinction-corrected fluxes of emission lines. $^\textrm{b}$In angstrom. $^\textrm{c}$Extinction coefficient. $^\textrm{d}$Equivalent width of hydrogen absorption lines in angstrom. $^\textrm{e}$Extinction-corrected flux, but not corrected for $f_\textrm{esc}$(LyC) in 10$^{-16}$ erg s$^{-1}$ cm$^{-2}$.}
\end{table*}

\begin{figure}
\centering
\includegraphics[angle=0,width=0.99\linewidth]{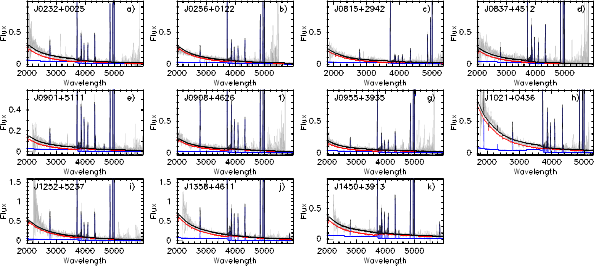}
\caption{SED fitting of the galaxy SDSS spectra. The rest-frame spectra, corrected for Milky Way and internal extinction, are shown by grey lines. The total, nebular, and stellar modelled intrinsic SEDs are shown by black, blue, and red lines, respectively. Fluxes and wavelengths are expressed in 10$^{-16}$ erg s$^{-1}$ cm$^{-2}$ \AA$^{-1}$ and angstrom, respectively.
\label{figa3}}

\vspace{0.5cm}

\centering
\includegraphics[angle=0,width=0.99\linewidth]{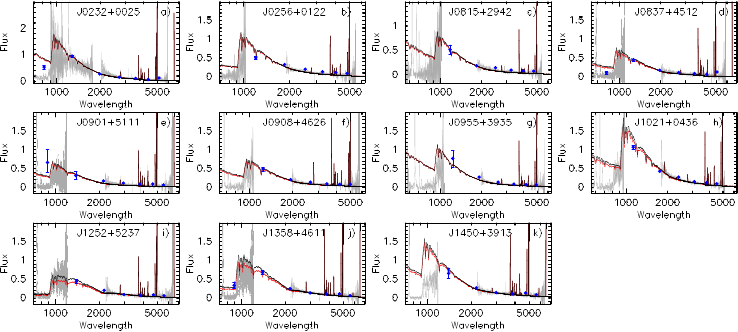}

\caption{Comparison of observed COS G140L and SDSS spectra (grey lines) and photometric data (blue-filled circles) with modelled SEDs. For the galaxy J0232+0025, the STIS G230L public spectrum from the program GO 17169 (PI: Schaerer) is added. \textit{GALEX} FUV and NUV fluxes with 1$\sigma$ error bars and SDSS fluxes in $u,g,r,i,z$ bands are shown by blue-filled circles. Modelled SEDs, which are reddened by the Milky Way extinction with $R(V)_\textrm{MW}$ = 3.1 and the internal extinction with $R(V)_\textrm{int}$ = 3.1 and 2.7, are shown by black and red solid lines, respectively. The emission lines in the LyC range of $<$912\AA\ correspond to the geocoronal Ly$\alpha$ $\lambda$1216\AA\ (brightest line) and the O~\textsc{i} $\lambda$1301\AA\ (weaker line). Fluxes are in 10$^{-16}$ erg s$^{-1}$ cm$^{-2}$\AA$^{-1}$, wavelengths are in angstrom. \label{figa4}}
\end{figure}

\end{appendix}

\end{document}